\documentclass[aps,prl,twocolumn,notitlepage,floatfix,nofootinbib]{revtex4-1}

\usepackage[pdftex]{graphicx}

\usepackage{amsmath,amssymb,amstext}
\usepackage{mathrsfs}
\usepackage{comment}
\usepackage{bbold}
\usepackage{dsfont}
\usepackage[T1]{fontenc}
\usepackage{color}
\newcommand{\ket}[1]{\left|#1\right>} 
\newcommand{\bra}[1]{\left<#1\right|} 
\newcommand{\id}{\mathbb{1}}  		

\usepackage{pdfsync}

\usepackage{hyperref}

\begin{document}

\title{Quantum-coherent mixtures of causal relations}

\author{Jean-Philippe W. MacLean$^{1,2,\ast}$, Katja Ried$^{1,2,3,\ast}$,  Robert W. Spekkens$^{3}$ and Kevin J. Resch$^{1,2}$}
\affiliation{$^1$Institute for Quantum Computing, University of Waterloo, Waterloo, Ontario, Canada, N2L 3G1}
\affiliation{$^2$Department of Physics \& Astronomy, University of Waterloo, Waterloo, Ontario, Canada, N2L 3G1}
\affiliation{$^3$Perimeter Institute for Theoretical Physics, Waterloo, Ontario, Canada, N2L 2Y5 \\
$^\ast$These authors contributed equally to this work.}

\begin{abstract}

 Understanding the causal influences that hold among parts of a system is
 critical both to explaining that system's natural behaviour and to controlling it
 through targeted interventions. In a quantum world, understanding causal relations
 is equally important, but the set of possibilities is far richer.  The two basic
 ways in which a pair of time-ordered quantum systems may be causally related are by
 a {cause-effect} mechanism  or by a {common cause} acting on both. Here, we show  
 a {coherent} mixture of these two possibilities.  We realize this nonclassical
 causal relation in a quantum optics experiment and derive a set of criteria for
 witnessing the coherence based on a quantum version of Berkson's effect, whereby two
 independent causes can become correlated upon observation of their common effect.  The
 interplay of causality and quantum theory lies at the heart of challenging
 foundational puzzles, including Bell's theorem and the search for quantum gravity.
\end{abstract}

\maketitle

\section*{Introduction}
 Unraveling the causal mechanisms that explain observed correlations is an important
 problem in any field that uses statistical data. For example, a positive correlation
 between the damage done by a fire and the number of fire fighters on scene does not
 imply that the firefighters caused the damage. Discovering causal relations has
 applications ranging from epidemiology and genetics to economics and policy
 analysis~\cite{Pearl_book,Spirtes_book}.  Causal explanation is also playing an
 increasingly important role in quantum physics.  It has recently gained prominence
 in the analysis of Bell's theorem and generalizations
 thereof~\cite{Fritz_2012,WoodSpekkens_2015, ChavesEtAl_2014, ChavesEtAl_2015}.
 Furthermore, causal structure is a close proxy for the structure of space-time in
 general relativity, and it has been suggested that we will have to abandon the
 notion of definite causal structure and instead allow superpositions thereof in
 order to develop a theory of quantum gravity~\cite{MarkopoulouSmolin_1998,
 Hardy_2007b}.  Understanding causality in a quantum world may provide new resources
 for future quantum technologies as we gain control over increasingly complex quantum
 systems. For instance, entanglement has been shown to provide a quantum advantage
 for causal inference in certain causal scenarios \cite{RiedEtAl_2015}.

 Classically, when two time-ordered variables are found to be statistically
 correlated, there are different causal mechanisms that could explain this. It could
 be that the early variable causally influences the later one, or that both are
 effects of a common cause.  Alternatively, the relation could be either cause-effect
 or common-cause with certain probabilities.  Most generally, there may be
 cause-effect and common-cause mechanisms acting simultaneously.  We refer to this as
 a {physical} (as opposed to probabilistic) mixture of the two mechanisms.  

 In a quantum world, there are additional possibilities.  Purely common-cause
 mechanisms are {intrinsically quantum} if they correspond to entangled bipartite
 states. Purely cause-effect mechanisms are intrinsically quantum if they correspond
 to channels that are not entanglement-breaking.  But quantum effects are not
 restricted to only these: as we will show, when common-cause and cause-effect
 mechanisms act simultaneously, one can have {quantum-coherent mixtures} of causal
 relations.   While conventional quantum mechanics can describe purely cause-effect
 and purely common-cause relations, quantum-coherent mixtures can only be represented
 using recent extensions of the formalism~\cite{Leifer_2006, AharonovEtAl_2009,
 ChiribellaEtAl_2009, Hardy_2012,Oreshkov2012,  Fitzsimons2013,Silva_2013}, in
 particular Refs.~\cite{LeiferSpekkens_2013, RiedEtAl_2015}.

 Chiribella~\cite{Chiribella_2012} and Oreshkov {\em et al.}~\cite{Oreshkov2012} have
 investigated coherent combinations of different causal {orderings}, specifically of
 $A$ causing $B$ and $B$ causing $A$. If realizable, such combinations would
 constitute a resource for  computational tasks~\cite{Hardy_2007}, with striking
 applications in gate
 discrimination~\cite{Chiribella_2012,AraujoEtAl_2014,ProcopioEtAl_2015}.  However,
 this possibility requires that $A$ and $B$ are not embeddable into a global causal
 order, whereas current physical theories implicitly assume such an ordering.  By
 contrast, we study a coherent combination of causal structures wherein $A$ is always
 temporally prior to $B$, a situation that is compatible with a global ordering and
 which therefore can be realized experimentally.   
 If the demonstrated applications of
 superpositions of causal orders noted above can be attributed to the novel
 possibilities that are allowed by quantum theory for combining causal relations, then other
 quantum-coherent mixtures of causal relations, such as cause-effect and
 common-cause, may also constitute a resource. 

 The present work provides a framework for describing the different ways in which
 causal relations may be combined and experimental schemes for realizing and
 detecting them.  
 We perform an experiment with photonic qubits that implements various such combinations 
 and observes their operational signatures.
Our main result is the experimental confirmation of the possibility of preparing a quantum-coherent mixture of common-cause and cause-effect relations.

\section*{Results}

\subsection{\small Signatures of different causal mixtures}

We seek to classify the causal relations that can hold between two quantum systems
and, in particular, to derive and detect an experimental signature of a quantum-coherent mixture of cause-effect and common-cause relations.  
The tools for this can be illustrated with a simpler example: a mixture of two cause-effect mechanisms in a scenario wherein two distinct causes influence a common effect.

 A signature for distinguishing physical from probabilistic mixtures can be derived
 from {Berkson's effect}, a phenomenon in  classical statistics whereby conditioning on a variable induces statistical correlations between its causal parents when they are otherwise uncorrelated. 
 Figure~1(a,b) provides an intuitive example.
 Note that the Berkson effect only arises if one has a combination of two causal
 mechanisms: in Fig.~1, 
 both teaching and research ability
 influence the hiring decision.  Crucially, the strength of the induced correlations
 can reveal how the two mechanisms are combined. In particular, probabilistic
 mixtures can only induce relatively weak correlations, as illustrated in
 Fig.~1(c)
 and proved rigorously in Supplementary Note 6. Correlations that are stronger than this bound bear witness to a physical mixture of causal mechanisms.

\begin{figure}
\center{\includegraphics{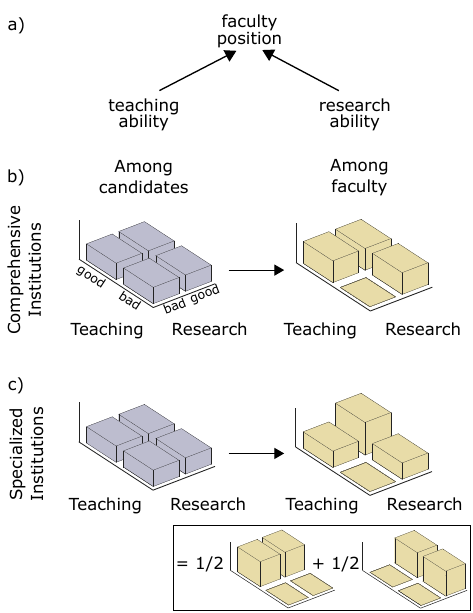} }
\caption{\footnotesize{
  {\bf An illustration of Berkson's effect when hiring faculty in different
  institutions.}
  (a) When applying for faculty positions, a candidate's success generally depends on
  their skills at both teaching and research. We assume that these abilities are
  statistically independent in the overall field of applicants. (b) At comprehensive
  institutions, the hiring process considers both skills and eliminates candidates who are
  both bad teachers and bad researchers. Consequently, the two abilities become
  {negatively correlated} among successful candidates.  (c) A set of specialized
  institutions, each one dedicated either purely to teaching or purely to research,
  select faculty based solely on the relevant ability in each case - a probabilistic
  mixture of both causal mechanisms as shown in the inset. Knowing that a candidate was
  successful in this scenario only reveals information about one of their abilities,
  and consequently induces weaker negative correlations than in (b), due to the
  larger fraction of faculty members who are skilled at both.
}}
\label{Berksonparadox}
\end{figure}

Quantum systems also exhibit the Berkson effect, with the strength of the induced correlations allowing one to distinguish physical from probabilistic mixtures. 
 However, the generalization to quantum systems adds a third category to this
 classification: post-selection may generate not just classical correlations, but
 {quantum} correlations (e.g., entanglement) between the causal parents.  We
 propose that this is a defining feature of a {quantum-coherent} mixture of
 causal mechanisms, a full definition of which will be developed below.

\subsection{ \small Causal relations between two classical systems} 

We now turn to the different causal relations between a pair of systems, labelled $A$ and $B$, with $A$ preceding $B$ in time.  
We begin by discussing the case of classical variables, which will motivate our
definitions for the quantum case. Figure~2(a) 
depicts the paradigm example, a drug trial, and Fig.~2(b-d)
introduces
useful representations of the possible causal relations.

\begin{figure}
 \centering
 \includegraphics[scale=0.8]{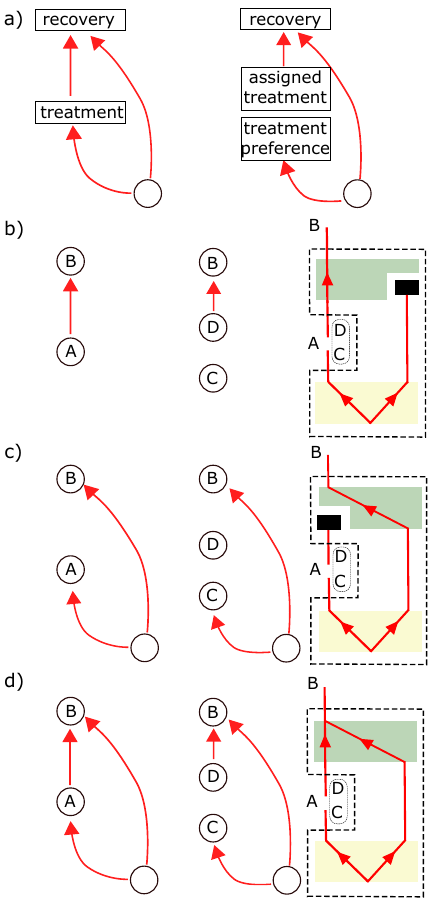} 
 \caption{\footnotesize{{\bf Causal relations between two time-ordered systems.} (a)
 A drug trial aims to discern whether treatment and recovery have a cause-effect
 relation, whether they share an unobserved common cause, or some combination of
 both. To this end, pharmaceutical companies randomly assign patients to take either
 the drug or a placebo, so as to evaluate the cause-effect relation. One may also
 track treatment preference in order to assess the common-cause relation. A complete
 characterization of the causal relation requires information about both versions of
 the treatment variable.  Abstract depictions of possible causal relations: (b)
 purely cause-effect, (c) purely common-cause and (d) general case, including
 mixtures of both mechanisms.  In directed acyclic graphs, arrows represent
 influences.  The variable $A$ is split into a pre-intervention version, denoted $C$,
 and a post-intervention version, denoted $D$.  The circuits realizing the causal
 relations, consisting of a preparation (yellow) of $A$ and an ancilla, followed by
 a coupling (green) between $A$ and the ancilla, which yields $B$.
}}
 \label{fig:DAGsAndCircuits}
 \end{figure}

Both the example of a randomized drug trial (Fig.~2a, right) 
and the circuit representation (Fig.~2d, middle) 
show that a complete description of the causal relation between $A$ and $B$ involves {two} versions of the variable $A$: the version prior to the randomizing intervention, denoted $C$ (treatment preference), has a purely common-cause relation to $B$, whereas the post-intervention version, $D$ (assigned treatment), directly influences $B$.
The causal relation between $A$ and $B$ is therefore completely specified by the stochastic map $P(CB|D)$. 

This scenario supports a more general version of the Berkson effect: 
conditioning on recovery can induce correlations not only between the assigned treatment and the unobserved common cause, but also, by extension, between assigned treatment and treatment preference.
These correlations bear witness to a combination of common-cause and cause-effect mechanisms, and their strength, as before, can distinguish different classes of combinations.

Before we develop a mathematical representation of these correlations in the quantum case, we first highlight a subtlety of the scenario by appealing to the classical case. 
In a randomized drug trial, the assigned treatment is controlled by the experimenter, hence there is no prior distribution over this variable. The object that encodes how assigned treatment correlates with treatment preference in the subpopulation that recovered is therefore not a joint distribution, but a {map} from assigned treatment to treatment preference. 
Given the overall stochastic map $P(CB|D)$, the subpopulation with $B=b$ is described by the element $P(C,B=b|D)$, which is a subnormalized stochastic map.
If one wishes to quantify the correlations encoded in this map using standard measures, one can use the following prescription to construct a joint distribution that is isomorphic to $P(C,B=b|D)$: let $u(D)$ denote the uniform distribution over $D$ and take $P^{b}(CD)\equiv P(C, B=b|D)u(D)/P_b$, where $P_b \equiv \sum_{CD} P(C, B=b|D)u(D)$ is a normalization factor. 
This object encodes the correlations we wish to study in a convenient form and, moreover, admits a close quantum analogue, as we will show. 
 
\subsection{ \small Causal relations between two quantum systems}

 If $A$ and $B$ are quantum systems, 
the input-output functionality of the circuits
in Fig.~2 
can be characterized using measurements on $B$ and an analogue of a randomized intervention on $A$, that is, a measurement followed by a random repreparation. As in the classical case, we split $A$ into $C$ and $D$.  
Mathematically, the circuit's functionality is represented by a trace-preserving, completely positive map from states on $D$ to states on the composite $CB$, $\mathcal{E}_{CB|D}: \mathcal{L}(\mathcal{H}_D) \to \mathcal{L}(\mathcal{H}_C\otimes \mathcal{H}_B)$ (where $\mathcal{L}(\mathcal{H}_X)$ denotes the linear operators over the Hilbert space of $X$), as can be inferred from Refs.~\cite{ChiribellaEtAl_2009,Hardy_2012,Oreshkov2012,Silva_2013,RiedEtAl_2015}, and which we term a {causal map}.

The Berkson effect on quantum systems is formalized as follows: consider a measurement on $B$, whose outcomes $b$ are associated with positive operators $\{ \Pi^b_B\}$. Finding an outcome $b$ implies correlations between $C$ and $D$, 
which are represented by a trace-non-increasing map from $D$ to $C$:
 $\mathcal{E}^b_{C|D} \equiv {\rm Tr}_B (\Pi^b_B \mathcal{E}_{CB|D})$ (analogous to the subnormalized stochastic map $ P(C,B=b|D)$).  Equivalently, we can represent this map using the quantum state $\tau^b_{CD}$ that one obtains by taking the operator that is Choi-isomorphic \cite{Choi1975} to $\mathcal{E}^b_{C|D}$ and normalizing it to have unit trace (analogous to the normalized distribution  $P^b(CD)$).  The correlations between $C$ and $D$ embodied in the map $\mathcal{E}^b_{C|D}$ can then be assessed using standard measures of correlation 
 on the state $\tau^b_{CD}$.
We say that the causal map exhibits a {Quantum Berkson Effect} if there exists a measurement $\{ \Pi^b_B\}$ such that for {every} outcome $b$, the induced correlations between $C$ and $D$, described by $\tau^b_{CD}$, are quantum.
For the purposes of this article, we take the presence of entanglement as a sufficient condition for quantumness.  Thus, our condition is that 
each $ \tau^b_{CD}$ be entangled, or equivalently, that each $\mathcal{E}^{b}_{C|D}$ be non-entanglement-breaking.
Using these definitions, we will now propose a classification of the possible causal relations between two quantum systems, as well as ways of distinguishing the classes. 

A causal map $\mathcal{E}_{CB|D}$ is {purely cause-effect} if it has the form
$\mathcal{E}_{CB|D}(\cdot) = \mathcal{E}_{B|D}(\cdot ) \otimes \rho_C$ (the analogue
of $P(CB|D)=P(B|D)P(C)$), which makes it compatible with the causal structure in
Fig.~2(b); 
and {purely common-cause} if $\mathcal{E}_{CB|D} (\cdot )= \rho_{CB} {\rm
Tr}_D(\cdot )$ (the analogue of $P(CB|D) =P(CB)$), which makes it compatible with the
causal structure in Fig.~2(c). 
 A causal map is said to be a {probabilistic mixture} of cause-effect and
 common-cause relations if there is a hidden classical control variable, $J$, which
 influences only $B$, such that for every value of $J$,  either $B$ depends only on
 $D$ or $B$ depends only on its common cause with $C$.  We show in Supplementary Note
 1 that every such causal map can be expressed as having just one term of each type,
 $\mathcal{E}_{CB|D} = w \mathcal{E} _{B|D}\otimes \rho_C +(1-w) \rho_{CB} \otimes
 {\rm Tr}_D$, where $0 \le w\le 1$ and ${\rm Tr}_B \rho_{CB}=\rho_C$.  The fact that
 the marginal on $C$ is the same in both terms follows from demanding that the
 control variable $J$ does not influence $C$.  This demand is justified by noting
 that a  probabilistic mixture of causal maps that are all purely cause-effect should
 also be purely cause-effect, but if the switch variable $J$ implementing this
 mixture could influence $C$ in addition to $B$, then it would itself constitute
 a common cause of $A$ and $B$.  If a causal map is {not} such a probabilistic
 mixture, then it is termed a {physical mixture} of cause-effect and common-cause
 mechanisms.

 Another distinction that is important for classifying causal relations between two
 quantum systems is whether the common-cause or cause-effect pathways of a given
 causal map are themselves quantum or not.  We propose that a sufficient condition for quantumness of
 the common-cause pathway is that there exists an orthogonal basis of states on $D$,
 indexed by $d$ and denoted $\rho_d$, such that the states on $CB$ induced by these
 preparations, $\tau^d_{CB} \equiv \mathcal{E}_{CB|D}(\rho_d)$, are {entangled}. 
 Similarly, a causal map is intrinsically quantum on the cause-effect pathway if
 there exists a measurement on $C$ that distinguishes a complete set of orthogonal
 states, indexed by $c$ and represented by projectors $ \Pi^c_C$, such that the
 induced correlations between $D$ and $B$ are quantum for every outcome $c$.  By the
 same reasoning established in the discussion of the Quantum Berkson Effect, these
 correlations are represented by trace-non-increasing maps from $D$ to $B$,
 $\mathcal{E}^c_{B|D} \equiv {\rm Tr}_C (\Pi^c_C \mathcal{E}_{CB|D})$, or
 equivalently by the normalized Choi-isomorphic states $\tau^c_{BD}$.  This allows us
 to propose a sufficient condition for quantumness in the cause-effect pathway that
 closely resembles the one for the common-cause pathway: the states $\tau^c_{BD}$
 must be entangled for all $c$.  

These distinctions give rise to eight classes of causal maps.  We here limit our attention to cases where the pathways are either both quantum or both classical, yielding four classes of interest, 
 illustrated in Fig.~3 and termed 
 {\sc ProbC}, {\sc PhysC}, {\sc ProbQ} and \textsc{PhysQ}.  
 The definition of the fifth class \textsc{(Coh)} is the central theoretical proposal
 of this article: a mixture of common-cause and cause-effect relations is {
 quantum-coherent} if the causal map is intrinsically quantum in both the
 common-cause and cause-effect pathways and it exhibits a Quantum Berkson
 Effect.  We note that the second requirement can only be satisfied if the causal map is
 a physical mixture, while the first implies that it is quantum in both
 pathways, hence \textsc{Coh} is contained in \textsc{PhysQ}.  We show in 
 Supplementary Note 2 that the inclusion is in fact strict.

\begin{figure}
 \centering
 \includegraphics[scale=1]{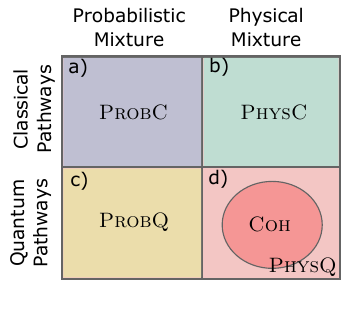} 
  \caption{\footnotesize{{\bf Classification of mixtures of causal relations between two quantum systems.}
One can distinguish whether the common-cause and cause-effect pathways are effectively classical or whether they are quantum, and whether they are combined in a probabilistic or a physical mixture. This gives rise to four categories of interest: a probabilistic mixture that is classical on both pathways ({\sc ProbC}), a physical mixture that is classical on both pathways {\sc(PhysC)}, a probabilistic mixture that is quantum on both pathways {\sc(ProbQ)}, and a physical mixture that is quantum on both pathways \textsc{(PhysQ)}. We leave aside cases wherein only one pathway is quantum.
The focus of this paper is the class \textsc{Coh}, which exhibits the Quantum Berkson Effect and therefore describes quantum-coherent mixtures of common-cause and cause-effect relations between $A$ and $B$.}}
\label{fig:classification}
 \end{figure}

\subsection{ \small Realizing \textsc{Coh} with a quantum circuit}

Figure~4 
presents quantum circuits that realize causal relations  between two qubits
exemplifying each of the classes.  Here, $E$ denotes the system that mediates between
$B$ and its common cause with $C$. System $F$ is introduced to make the gate
$\mathcal{E}_{BF|DE}$ preserve dimensionality, but it is discarded afterwards.
The initial state $\rho_{CE}$ in all cases is the maximally entangled state
\begin{equation}
\ket{\Phi^+}  \equiv\frac{1}{\sqrt{2}}(\ket{HH}+\ket{VV}),
\label{eq:phiplus}
\end{equation}
where $\ket{H}$, $\ket{V}$ denote the eigenstates of the Pauli operator $\sigma_z$,
anticipating the identification as horizontal and vertical polarization states of our photonic qubits. The gate $\mathcal{E}_{BF|DE}$ that
realizes a coherent mixture 
applies the partial swap unitary,
  \begin{align}
    \begin{split}
    U_{BF|DE}=
   \tfrac{1}{\sqrt{2}}  \mathds{1}_{B|D} \otimes \mathds{1}_{F|E}
    + i \tfrac{1}{\sqrt{2}}  \mathds{1}_{B|E} \otimes \mathds{1}_{F|D},
  \end{split}\label{eq:partialswap}
  \end{align}
 where $\mathds{1}_{Y|X}$ denotes the identity operator from $X$ to $Y$.  This unitary coherently combines the two-qubit identity operator, $\mathds{1}_{B|D} \otimes \mathds{1}_{F|E}$, which realizes a purely cause-effect relation between $A$ and $B$, and the swap operator, $ \mathds{1}_{B|E} \otimes \mathds{1}_{F|D}$, which realizes a purely common-cause relation. 
 
 \begin{figure*}
 \centering
 \includegraphics[scale=0.8]{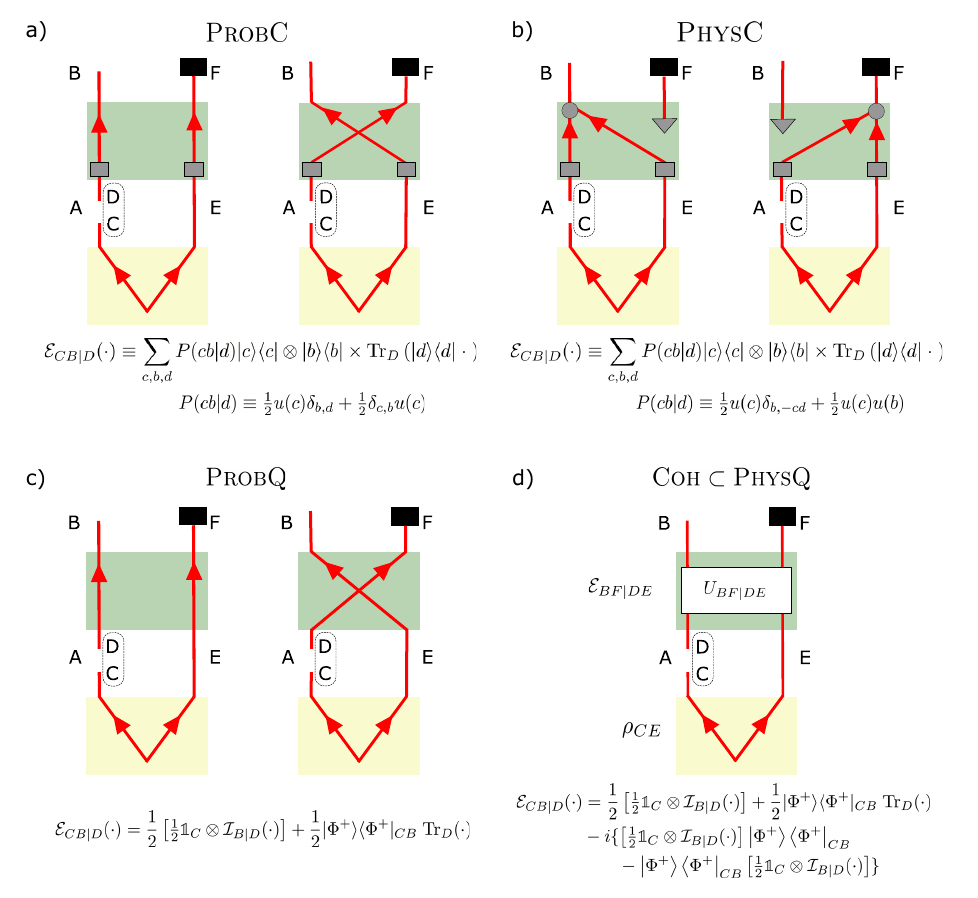} 
    \caption{\footnotesize{{\bf Quantum circuits which realize examples of
    different classes of causal relations between two qubits.} { 
    Circuits for  four combinations of cause-effect and common-cause mechanisms: {\bf
    a)} \textsc{ProbC}, {\bf b)} \textsc{PhysC}, {\bf d)} \textsc{ProbQ}, {\bf d)}
    \textsc{Coh}$\subset$\textsc{PhysQ}.}  In all circuits, $C$ and $E$ are initially
    prepared in the maximally entangled state $\ket{\Phi^+}$ and $F$ is discarded at
    the end. The examples differ only in the choice of the gate
    $\mathcal{E}_{BF|DE}$, as described in the text. Panels with two circuits
    represent an equal probabilistic mixture of both scenarios.  Black squares
    represent the trace operation, grey squares represent complete dephasing
    operations, grey triangles represent preparations of the completely mixed state,
    grey circles represent the classical XNOR gate, which sets $b=-ed$ for binary
    variables $b,e,d$ taking the values $\{\pm1\}$, and the two-qubit unitary
    $U_{BF|DE}$ is the partial swap given by equation~(\ref{eq:partialswap}).  
    Below each example, we specify the causal map $\mathcal{E}_{CB|D}$ obtained from the
    state $\rho_{CE}$ and the gate $\mathcal{E}_{BF|DE}$ via $\mathcal{E}_{CB|D}
    (\cdot) = {\rm Tr}_F \circ \mathcal{E}_{BF|DE} (\;\cdot \; \otimes \rho_{CE} )$.
    Lowercase letters $c,b,d$ represent classical binary variables, $P(cb|d)$
    represents a conditional probability distribution over these, $\delta_{x,y}$
    denotes the Kronecker delta function, and $u(x)$ denotes the uniform distribution
  over $x$.
   }
    \label{fig:example circuits} }
 \label{fig:ExampleCircuits} \end{figure*}

Combining the circuit elements $\rho_{CE}$ and $\mathcal{E}_{BF|DE}$ and tracing out $F$, we find
\begin{align}
\label{eq:Ecoh}
\begin{split}
  & \mathcal{E}_{CB|D}(\cdot)=
  \frac{1}{2} \left[ \tfrac{1}{2}\id_C \otimes\mathcal{I}_{B|D}(\cdot) \right]
  + \frac{1}{2}\ket{\Phi^+}\bra{\Phi^+}_{CB} {\rm Tr}_D(\cdot)\\
  &\hspace{1.4cm}-i\Big\{ \left[\tfrac{1}{2}\id_C \otimes
  \mathcal{I}_{B|D}(\cdot)\right]\ket{\Phi^+}\bra{\Phi^+}_{CB}\\\nonumber &\hspace{2.1cm}
  -\ket{\Phi^+}\bra{\Phi^+}_{CB}[\tfrac{1}{2}\id_C\otimes
  \mathcal{I}_{B|D}(\cdot)]\Big\}.
\end{split}
\end{align}
The first term applies the identity channel from $D$ to $B$, $\mathcal{I}_{B|D}$, 
whereas the second prepares $C$ and $B$ in the maximally entangled state
$\ket{\Phi^+}$. The cross terms encode coherences between these two causal relations.
One can verify that this causal map is quantum in both the cause-effect and the
common-cause pathway. It also exhibits a Quantum Berkson Effect: if $B$ is measured
to be in the state $\ket{H}$, then
 \begin{align}
   \tau_{CD}^{H}=&\frac{1}{2}
   \ket{HH}\bra{HH}+\frac{1}{2}\ket{\varphi}\bra{\varphi}
 \end{align}
where $\ket{\varphi} \equiv  \tfrac{1}{\sqrt{2}} (\ket{HV}-i \ket{VH}$), hence $\tau^H_{CD}$ is
entangled.  If instead $B$ is measured to be in the state $\ket{V}$, $\tau^V_{CD}$ is
similarly entangled.
The causal map therefore belongs to \textsc{Coh}.

 The example of \textsc{ProbQ} from Fig.~4(c) 
 obtained by replacing the partial swap with an equal probabilistic mixture of
 identity and swap, which eliminates the cross terms in equation~(\ref{eq:Ecoh}). The
 result is a manifestly probabilistic mixture of causal relations that is
 nevertheless quantum on both pathways. One can further modify this gate to realize
 the example of \textsc{ProbC} in Fig.~4(a) 
 complete dephasing operations on its inputs, $D$ and $E$, which effectively reduces
 both qubits to classical bits.

 The example of \textsc{PhysC} presented in Fig.~4(b) 
 also begins by complete dephasing on $D$ and $E$ to ensure that both pathways are
 indeed classical.  The simplest example of a physical mixture would then be one
 wherein $B$ is a nontrivial function of both $D$ and $E$. However, since we wish to
 realize all of these examples with a single experimental set-up, we consider instead
 a probabilistic mixture of two gates, one of which has $B$ as a nontrivial function
 of both $D$ and $E$, whereas the other prepares $B$ in the completely mixed state.
 The expression $\mathcal{E}_{CB|D}$ for each example and the proof that they are all
 indeed representatives of their classes are provided in the Supplementary Note 2. 

\subsection{ \small Experimental signatures of causal relations}
\begin{figure*}
  \centering
    \label{fig:setup} 
    \includegraphics[scale=0.9]{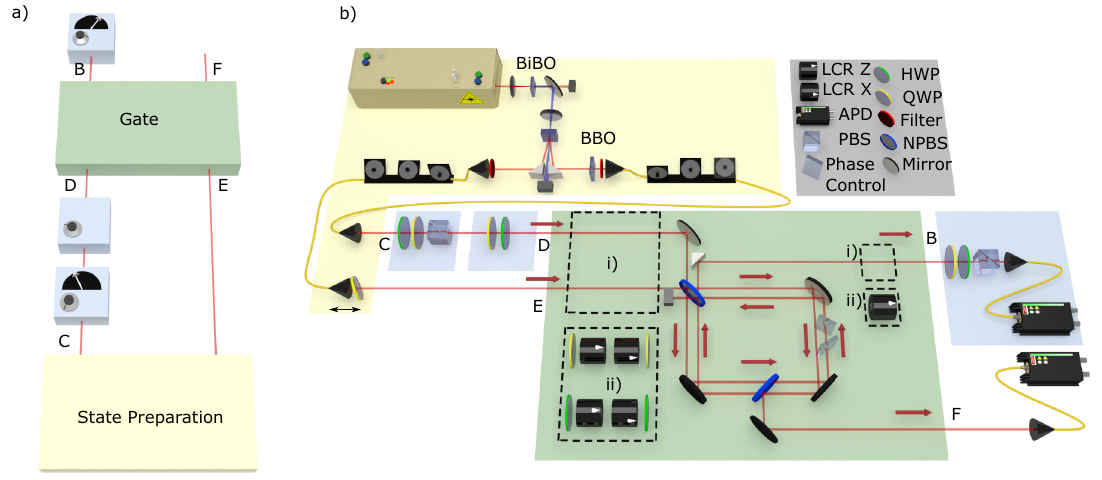} 
    \caption{\footnotesize{ {\bf Optical implementation of different causal
    relations.} {\bf a)} Schematic diagram of the experiment. The gate allows the
    implementation of different ways of combining a common-cause relation (between
    $C$ and $B$) with a cause-effect relation (between $D$ and $B$).  The initial
    preparation targets the maximally entangled state $\ket{\Phi^+}$. One photon is
    measured at $C$ and reprepared at $D$; then both are sent through the gate. The
    photon at $B$ is detected in coincidence with the photon at $F$ to post-select
    only on those cases wherein a pair was produced. {\bf b)} Experimental setup,
    including the polarization entangled photon source and the partial swap, which
    implements the unitary $ U_{BF|DE}= \tfrac{1}{\sqrt{2}}  \mathds{1}_{B|D} \otimes
    \mathds{1}_{F|E} + i \tfrac{1}{\sqrt{2}}  \mathds{1}_{B|E} \otimes
    \mathds{1}_{F|D}$ by tilting the glass plates to a specific angle in the Sagnac
    interferometer. {\bf i)} For the quantum mixtures, no dephasing is applied and
    a translation stage adjusts the delay of the photon at $E$ with respect
    to the one at $D$. {\bf ii)} For the classical mixtures, the LCRs and wave plates
    are used to apply complete dephasing on $D$, $E$, and $B$, respectively. Notation
    for optical elements: Bismuth-Borate (BiBO),  $\beta$-Barium-Borate (BBO),
    half-wave plate (HWP), quarter-wave plate (QWP), liquid-crystal retarder (LCR),
    polarizing beam splitter (PBS), non-polarizing beam splitter (NPBS), avalanche
  photo diode (APD).}}
\end{figure*}

 The four circuits of Fig.~4 
 are experimentally realized using the set-up of Fig.~5. 
 The polarization degrees of freedom of different photon modes constitute the qubits
 in our circuit.  We use downconversion to prepare the photonic modes $C$ and $E$ in
 the maximally entangled polarization state $\ket{\Phi^+}$.

 To realize our example of \textsc{Coh} (Fig.~4(d)), 
 the partial swap in
 equation~(\ref{eq:partialswap}) is implemented using linear optics~\cite{CernochEtAl_2008}.
 Here, we significantly improve the stability of the experimental concept of Ref.~\cite{CernochEtAl_2008}
 by incorporating the working principle around a displaced Sagnac interferometer. The
 other three examples from Fig.~4 
 are obtained by variations on this set-up.  Delaying the photon in mode
 $E$ relative to the one in mode $D$ prevents two-photon interference at the first
 beam splitter of the Sagnac interferometer, so that the interferometer implements a probabilistic mixture of identity and swap operations (our example of \textsc{ProbQ},
 Fig.~4(c)). 
 This latter circuit realizes the same causal map implemented in Ref.~3, which focused on the task of 
resolving probabilistic mixtures of cause-effect and common-cause relations. 
 However, the experimental setup of Ref.~3 
 could not realize physical mixtures, which are 
 required to address the broader question, investigated in the present work, of how
 these two extremes may be combined in general.

 Both causal pathways can be made classical by passing the modes through completely dephasing
 channels on $D$, $E$ and $B$. The example of \textsc{ProbC} from
 Fig.~4(a) 
 is realized by dephasing in the $\{\ket{H},\ket{V}\}$ basis on all three. The
 example of \textsc{PhysC} from Fig.~4(b) 
 is also achieved by implementing complete dephasing, but in different bases:
 $\{\ket{R},\ket{L}\}$ on $D$, $\{\ket{D},\ket{A}\}$ on $E$ and $\{\ket{H},\ket{V}\}$
 on $B$. 
 For further details on how to implement the four example classes
 of causal structures using a single experimental setup, see Supplementary Note 3.
  
 We characterize the causal maps realized in the experiment using
 tomography~\cite{RiedEtAl_2015,ChiribellaEtAl_2009,bisio_quantum_2011,Silva_2013}:
 measurements on $C$ and $B$ and preparations on
 $D$, each ranging over the six eigenstates of Pauli observables, 
 allow us to reconstruct the map using a least-squares fit. The causal maps obtained
 from the four circuits in Fig.~4 
 are shown in Fig.~6 
 and achieve fidelities above $93\%$ with their respective targets.
 Although these maps encode a complete description of the causal relation realized
 between $A$ and $B$, since our goal is only to classify the causal relation, we will
 introduce and evaluate specific indicators that can achieve this purpose with fewer
 measurements and preparations.

 \begin{figure*} \centering
   \includegraphics[width=1\textwidth]{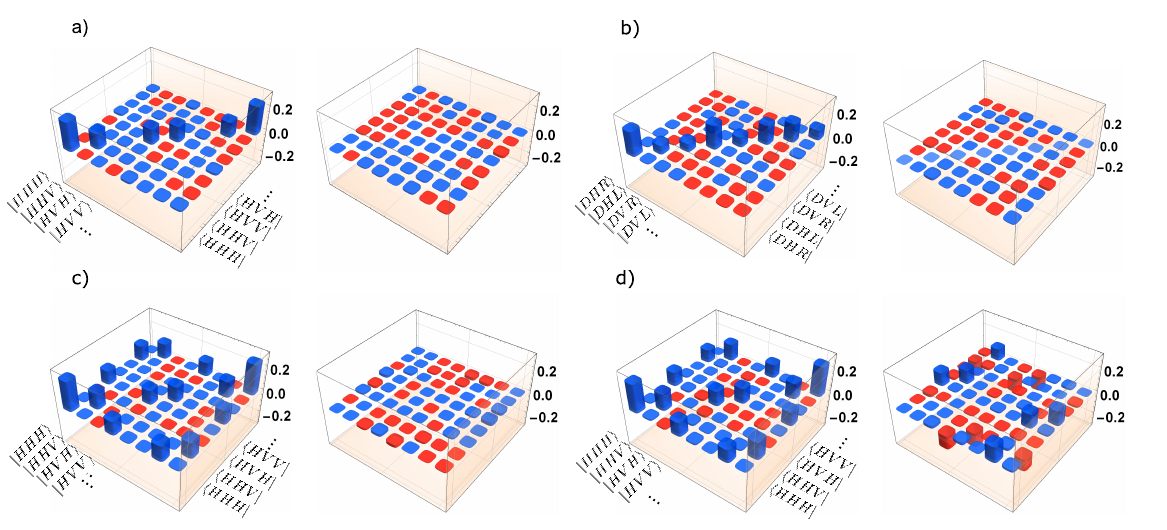}
   \caption{\footnotesize{{\bf Reconstructed causal maps.} Reconstruction of the real
   and imaginary parts of the Choi state $\tau_{CBD}={\rm Tr}_{FD'}
   \left[(\mathcal{E}_{BF|D'E}\otimes\mathcal{I}_{CD})
   \left(\ket{\Phi^+}\bra{\Phi^+}_{D'D}\otimes\rho_{CE}\right)\right]$ when targeting
   four combinations of cause-effect and common-cause mechanisms:  {\bf a)}
   \textsc{ProbC}, {\bf b)} \textsc{PhysC}, {\bf d)} \textsc{ProbQ}, {\bf d)}
   \textsc{Coh}.  The causal maps for (a) and (b) are given in the local bases that
   diagonalize the target map, in order to make explicit their classical nature; no
   such bases exist for (c) and (d).  Blue (red) colour bars represent positive
   (negative) values.  The fidelities\cite{jozsa_fidelity_1994}, $F\equiv\left[{\rm
   Tr}\sqrt{\tau^\frac{1}{2}\tau_{\rm th}\tau^\frac{1}{2}}\right]^2$, to the
   theoretically calculated Choi state $\tau_{\rm th}$ are high, at $(98.1 \pm
   0.2)\%$, $(98.06 \pm 0.08)\%$, $(97.1 \pm 0.1)\%$ and $(93.7 \pm 0.3)\%$,
   respectively, verifying that the experiment is performing as intended.
   Uncertainties on fidelities indiciate one standard deviation and are estimated
   using Monte-Carlo simulations with Poissonian noise on photon counts.  Notation
   for polarization states: 
   $\ket{H}$, horizontal, $\ket{V}$ vertical,
   $\ket{D}=1/\sqrt{2}(\ket{H}+\ket{V})$ diagonal,
   $\ket{A}=1/\sqrt{2}(\ket{H}-\ket{V})$ anti-diagonal,
   $\ket{R}=1/\sqrt{2}(\ket{H}+i\ket{V}$ right-circular, 
   $\ket{L}=1/\sqrt{2}(\ket{H}-i\ket{V})$ left-circular.}}
 \label{fig:causal maps} \end{figure*}

A witness of physical mixture (as opposed to probabilistic) can be evaluated using only 
 measurements of the Pauli observables $\sigma_x$ on $C$ and $\sigma_z$ on $B$, with
 outcomes $c,b=\pm1$, while preparing the $d$ eigenstate of $\sigma_y$ on $D$, with
 $P(d=\pm1)=\frac{1}{2}$. (Different choices of Pauli observables generate a family
 of such witnesses.) For subsets of this data with different values of $b$, one can compute
 the covariance of $c$ and $d$, which we denote ${\rm cov}(c,d|b)$ (see Supplementary Note 5 for details). 
Letting $P(b)$ denote the probability of obtaining the outcome $b$, we define our witness to be
\begin{equation}
  \mathcal{C}_{CD}\equiv 2\sum_b b P(b)^2 {\rm cov}(cd|b).
\end{equation}
We show  in  Supplementary Note 5  that $\mathcal{C}_{CD}=0$ for all probabilistic mixtures of common-cause and cause-effect, which implies that $\mathcal{C}_{CD}\neq0$ heralds a physical mixture.

We reconstruct the operators $\tau^c_{BD}$, $\tau^d_{CB}$ and $\tau^b_{CD}$ 
using subsets of tomographic data (for example, using only runs that found $B$ in the state $\ket{H}$ to reconstruct $\tau^H_{CD}$). The entanglement of these states is quantified by the negativity~\cite{VidalWerner_2002},
  \begin{equation}
    \mathcal{N}^z_{XY}\equiv\frac{1}{2}({\rm Tr}|T_Y(\tau^z_{XY})|-1),
    \label{eq:negativity} 
  \end{equation} 
 where $T_Y(\cdot)$ denotes transposition on $Y$. Quantumness in the cause-effect and
 common-cause pathways is therefore witnessed by $\mathcal{N}^c_{BD}>0\;\forall c$
 and $\mathcal{N}^d_{CB}>0\;\forall d$, respectively, while
 $\mathcal{N}^b_{CD}>0\;\forall b$ witnesses a Quantum Berkson Effect. 
 See Supplementary Note 4 for more details on obtaining the negativity of the pre-
 and post-selected states from experimental data.

 \begin{figure} 
   \centering
   \includegraphics[scale=1]{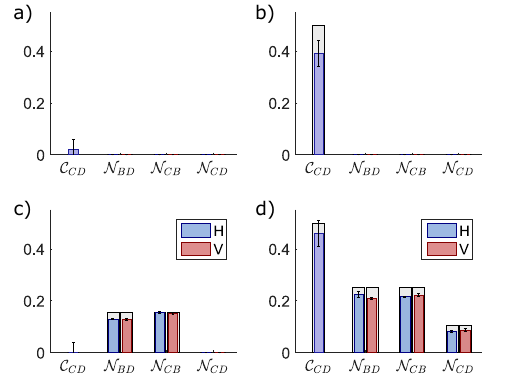}
   \caption{\footnotesize{{\bf Classifying causal relations using induced
   correlations.} 
   For each circuit in Fig.~4, 
   we present theoretical (grey) and experimental (coloured) values for the different
   witnesses of causal relations. Circuits (b) and (d) have $\mathcal{C}_{CD}\neq0$,
   witnessing physical mixtures, whereas (a) and (c) are consistent with
   probabilistic mixtures, since $\mathcal{C}_{CD}$ is zero within one standard
   deviation.  Circuits (c) and (d) show evidence of intrinsically quantum
   cause-effect and common cause mechanisms, $\mathcal{N}^c_{BD}\neq 0$ and
   $\mathcal{N}^d_{CB}\neq 0$ for $d,c=H,V$, whereas (a) and (b) are consistent with
   classical mechanisms.  Only (d) has $\mathcal{N}^b_{CD}\neq0$ for $b=H,V$,
   witnessing a Quantum Berkson Effect.  Uncertainties indicate one standard
   deviation and are estimated using Monte Carlo simulations, assuming Poissonian
   noise on the photon counts.
 }}  \label{fig:results}
 \end{figure}
 Figure~7 
 summarizes the values of these witnesses for the four circuits of Fig.~4 
 along with the corresponding theoretical expectations. The indicators
 $\mathcal{N}^c_{BD}$, $\mathcal{N}^d_{CB}$ and  $\mathcal{N}^b_{CD}$ are evaluated
using tomographic reconstructions of the operators $\tau^c_{BD}$, $\tau^d_{CB}$ and $\tau^b_{CD}$,
 respectively, under preparations (on $D$) and measurements (on $C$ and $B$) of
 $\{\ket{H},\ket{V}\}$.
 Scenarios (b) and (d) show evidence of a physical mixture, with
 $\mathcal{C}_{CD}=0.40 \pm 0.02$ and $\mathcal{C}_{CD}=0.46 \pm 0.02$. while
 scenarios (c) and (d) exhibit quantumness in the common-cause and cause-effect
 pathways. 
 We find $\mathcal{N}^d_{CB}$ and $\mathcal{N}^c_{BD}$ non-zero for
 $c,d\in \{H,V\}$. These signatures confirm that we realized physical
 mixtures and quantum common-cause and cause-effect mechanisms as intended. 

The most important indicator for our purposes is $\mathcal{N}^b_{CD}$, which verifies the Quantum Berkson Effect. 
As expected, scenarios (a-c) have $\mathcal{N}^b_{CD}=0$ (to within statistical error) and are therefore compatible with an {incoherent} mixture of common-cause and cause-effect relations. Scenario (d), however, exhibits Berkson-type induced entanglement, with $\mathcal{N}^H_{CD}=(0.083 \pm 0.003)$ and $\mathcal{N}^V_{CD}=(0.087 \pm 0.004)$ 
This, combined with the evidence of quantumness of each individual mechanism,
constitutes a clear signature of the class \textsc{Coh}, a quantum-coherent mixture of cause-effect and
common-cause relations.

\section*{Discussion}

 A priori, it is not obvious how one ought to define a quantum-coherent combination
 of causal relations, in particular a quantum-coherent combination of common-cause
 and cause-effect relations.  In this article, we have proposed a particular
 definition and demonstrated the possibility of realizing such quantum-coherence
 experimentally.  There are two distinct notions of quantum coherence that one might
 think are pertinent to our problem, and both feature in our definition. 

 The first notion of quantum coherence applies to elements of a set of alternatives
 that are jointly exhaustive and mutually exclusive, such as the eigenstates of some
 observable.  In this case, an incoherent mixture is a probabilistic mixture of the
 alternatives and consequently it is reasonable to define a state as exhibiting
 coherence whenever it cannot be expressed as such a probabilistic mixture.  However,
 the sorts of causal relations that we here seek to combine coherently,
 a cause-effect relation and a common-cause relation, {do not} constitute mutually
 exclusive alternatives.  A pair of systems may be connected by {both} a cause-effect
 relation and a common cause.  The possibility of two causal mechanisms acting
 simultaneously necessitates the category of {physical mixtures} of causal
 mechanisms.  Given that such physical mixtures can arise classically, the mere
 inapplicability of a probabilistic mixture should not lead one to infer the presence
 of quantumness.   This is why we use additional criteria for judging a combination
 of causal relations to be a quantum-coherent combination.  Because physical mixtures
 are distinguished by the strength of the induced correlations in the Berkson effect,
 we have proposed that a necessary condition for having a quantum-coherent mixture is
 that the Berkson-induced correlations exhibit entanglement. 

 The second notion of quantum coherence is the one relative to which different
 {systems} are said to be coherent with one another: for independent systems, this
 occurs when their joint state is entangled, for the input and output of a quantum
 channel, this occurs when the channel is not entanglement-breaking.  In the causal
 context, therefore, a common-cause relation between a pair of systems can be judged
 coherent if the state of the systems is entangled, while a cause-effect relation
 between a pair of systems can be judged coherent if the associated channel is not
 entanglement breaking.  This second notion of coherence is applicable, therefore, to
 individual causal pathways rather than the manner in which they are combined.
 Consequently, we have proposed that another necessary condition for a mixture of
 cause-effect and common-cause relations to be quantum-coherent is that each of the
 pathways, common-cause and cause-effect, are themselves coherent. 

 Our approach to defining quantum-coherent combinations of different causal relations
 differs significantly from the one suggested in recent work seeking to define
 superpositions of different causal orders.  The proposal for witnessing  causal
 nonseparability in Ref.~\cite{AraujoEtAl_2015}, for instance, judges the causal
 order between a pair of systems to be quantum-indefinite whenever the causal map
 cannot be written as a probabilistic mixture of terms with definite causal orders.
 However, from our perspective, $A$ causing $B$ is not necessarily mutually exclusive
 to $B$ causing $A$ (just as a common-cause relation is not mutually exclusive to
 a cause-effect relation).  To imagine both acting simultaneously -- which we would
 term a {physical mixture} of the two cause-effect relations -- is simply to imagine
 the possibility of causal cycles.  This is an exotic possibility, but one that is
 classically meaningful.  As such, having a causal map that is not a probabilistic
 mixture of causal orders need not, by itself, be evidence of quantumness.
 See Supplementary Discussion 1 for more on the related topic of superposition of
 causal orders.

 As we progress to studying more complex scenarios, for instance, involving a larger
 number of systems, we are likely to find an even wider range of types of coherence,
 resembling the many types of entanglement that arise when more than two parties are
 involved.  The problem of classifying the causal possibilities in this case---in
 particular quantum-coherent mixtures of various relations---is significantly more
 complex than the one considered here.  Developing such a classification for an
 arbitrary number of systems with arbitrary dimensionality constitutes an important
 pillar in the new research programme that seeks to understand causality in quantum
 theory.

 The understanding of new, uniquely quantum, combinations of common-cause and
 cause-effect relations introduced in this article also has implications for the
 topic of non-Markovianity. In the context of the dynamics of open quantum systems,
 the assumption of Markovianity states that the environment with which the principal
 system interacts has no memory and hence is unable to preserve a record of earlier
 states of the system. (See Ref.~\cite{RivasEtAl_2014,VacchiniEtAl_2011} for a review
 of proposed quantum statements of Markovianity.) This assumption simplifies the
 mathematical treatment of the system's dynamics considerably, but in most realistic
 models it holds only approximately, which has sparked considerable interest in
 quantum non-Markovianity in recent years.  From the perspective of causal modeling,
 non-Markovianity arises when the environment acts as a common cause of the system at
 different times (in addition to the cause-effect relations arising from the
 evolution of the system itself).  As such, the fact that there are intrinsically
 quantum ways of mixing common-cause and cause-effect relations implies a greater
 variety of types of non-Markovianity than one sees classically.

\section*{Methods}

  \subsection{\small Photon Source}
  We produce polarization entangled photon pairs using spontaneous parametric
  downconversion in two type-I nonlinear crystals. We begin with a Ti:Sapphire laser,
  centred at 790 nm with a spectral bandwidth of 10.5nm, a repetition rate of 80 MHz,
  and an average power of 2.65 W. The laser light is frequency doubled in a 2-mm thick
  bismuth-borate (BiBO) crystal, which creates a pump beam of 0.65 W centred at
  395nm with a 1 nm FWHM bandwidth.  With two cylindrical lenses, the
  pump is focused onto a pair of 1~mm $\beta$-barium-borate (BBO) crystals with
  orthogonal orientations for type-I spontaneous parametric downconversion. Bandpass
  filters are placed to reduce background noise from the pump. Additional
  compensation crystals are used in order to counteract the effects of temporal
  and spatial walkoff \cite{lavoie_experimental_2009}.  Polarization entangled photon
  pairs at 790 nm are prepared in the state $\ket{\Phi^+}$, which we achieve with
  $(96.31 \pm 0.08)\%$ fidelity. Inteference filters on both sides set the photon
  bandwidths to 3nm. The photons are then coupled into single mode fibres and sent
  towards the partial swap. The polarization is set with polarization controllers and
  the phase of the entangled state is tuned by tilting a quarter-wave plate (QWP) at
  the output of one of the fibres.
  
  \subsection{\small Implementing the partial swap} The partial swap uses
  a folded displaced Sagnac interferometer, with two 50/50 beam splitters and two
  NBK-7 glass windows, which are counter rotated in order to set the phase with
  minimal beam deflection.  The visibility of the Sagnac interferometer without
  background subtraction is $\left(93.6\pm0.1 \right)\%$. This is measured by
  blocking one input ($D$ or $E$) to the gate and measuring the number of photons at
  the output $B$ as a function of the window angles in the Sagnac interferometer.
  For the coherent partial swap to be effective, photon pairs in modes $D$ and $E$
  must undergo two-photon quantum interference on a beam splitter prior to entering
  the gate.  Hong-Ou-Mandel (HOM) interference between photons input at $D$ and $E$
  is measured at the first beam splitter using a translation stage on input $E$.
  A dip in visibility of $(95\pm2)\%$ is achieved. To implement the gate for the
  class \textsc{ProbQ}, a delay of 3 ps is added to photon $E$ which removes the
  interference. For all other cases, the delay is set to a value corresponding to the
  centre of the HOM dip to maximize the two photon interference.

  \subsection{\small Dephasing Channels}
  The dephasing channels before and after the Sagnac interferometer are implemented
  using variable liquid crystal retarders (LCR), which exhibit a voltage dependent
  birefringence, introducing a relative phase of $0$ or $\pi$ on orthogonal
  polarization states. Probabilistic dephasing is achieved by switching them on and
  off at random at a rate of 10 Hz, with probability $1/2$ of implementing a $\pi$
  phase shift during each interval.  The input and outputs can be dephased on
  a different polarization bases.  Dephasing along the $\{\ket{D},\ket{A}\}$ basis is
  achieved with the LCR axis at $0^\circ$, and in the $\{\ket{H},\ket{V}\}$ basis at
  $45^\circ$. Two quarter-wave plates on either side of the LCR after $D$ and two
  half-wave plates on either side of the LCR after $E$ are used to rotate between the
  different dephasing bases required for the classical mixtures.

  \subsection{\small Measurement Procedure}
  The experiment proceeds in the following way. The unitary $U_{BF|DE}$ is set by
  adjusting the window angles in the Sagnac interferometer such
  that the phase difference between the two paths is $\pi/2$. A HOM dip is then
  measured and the arrival time of the photons is set with a translation stage at $E$.
  The entangled state on $C$ and $E$ is initialized by preparing and measuring remote
  entanglement between $C$ and $B$.  The polarization is measured using a half-wave
  plate (HWP), quarter-wave plate (QWP), and polarizing beam splitter (PBS) in
  sequence. The HWP and QWP are adjusted so that one polarization state can pass
  through the PBS.  The polarization at $C$ is measured with a HWP and QWP, and
  assuming the photon has passed through the PBS, the polarization is reprepared with
  another QWP and HWP at $D$.  Photons are then sent to the partial swap gate, and
  the polarization at $B$ is measured using another QWP and HWP.  Coincidence counts
  at $F$ and $B$ are measured using Silicon avalanche photodiodes and a coincidence
  logic with a coincidence window of 3 ns.  Coincidences are measured at a rate of
  approximately 1 kHz.  We measure the different combinations of polarization
  eigenstates and repeat the procedure for the four different causal scenarios.

\textbf{Acknowledgments} 
   This research was supported in part by the Foundational Questions Institute
   (grant number FQXI-RFP-1516), the Natural Sciences and Engineering Research
   Council of Canada (NSERC), Canada Research Chairs, Industry Canada and the Canada
   Foundation for Innovation (CFI).  Research at Perimeter Institute is supported by
   the Government of Canada through Industry Canada and by the Province of Ontario
   through the Ministry of Research and Innovation.

 \textbf{Contributions}
   RWS, KR and KJR conceived the original idea for the project.  
   KR and RWS developed the project and the theory.  
   JPWM and KJR designed the experiment.
   JPWM performed the experiment and the numerical calculations.  
   JPWM, KR, RWS and KJR analyzed the results.  
   JPWM and KR wrote the first draft of the paper and all authors contributed to the final version.

  \textbf{Corresponding author}
   Correspondence and request for materials should be
   addressed to J.P.W.M. (jpmaclean@uwaterloo.ca) or K.R.  (katja.ried@uibk.ac.at).

\bibliography{../CoherentCausation,../causalityexp}

\begin{thebibliography}{32}%
\makeatletter
\providecommand \@ifxundefined [1]{%
 \@ifx{#1\undefined}
}%
\providecommand \@ifnum [1]{%
 \ifnum #1\expandafter \@firstoftwo
 \else \expandafter \@secondoftwo
 \fi
}%
\providecommand \@ifx [1]{%
 \ifx #1\expandafter \@firstoftwo
 \else \expandafter \@secondoftwo
 \fi
}%
\providecommand \natexlab [1]{#1}%
\providecommand \enquote  [1]{``#1''}%
\providecommand \bibnamefont  [1]{#1}%
\providecommand \bibfnamefont [1]{#1}%
\providecommand \citenamefont [1]{#1}%
\providecommand \href@noop [0]{\@secondoftwo}%
\providecommand \href [0]{\begingroup \@sanitize@url \@href}%
\providecommand \@href[1]{\@@startlink{#1}\@@href}%
\providecommand \@@href[1]{\endgroup#1\@@endlink}%
\providecommand \@sanitize@url [0]{\catcode `\\12\catcode `\$12\catcode
  `\&12\catcode `\#12\catcode `\^12\catcode `\_12\catcode `\%12\relax}%
\providecommand \@@startlink[1]{}%
\providecommand \@@endlink[0]{}%
\providecommand \url  [0]{\begingroup\@sanitize@url \@url }%
\providecommand \@url [1]{\endgroup\@href {#1}{\urlprefix }}%
\providecommand \urlprefix  [0]{URL }%
\providecommand \Eprint [0]{\href }%
\providecommand \doibase [0]{http://dx.doi.org/}%
\providecommand \selectlanguage [0]{\@gobble}%
\providecommand \bibinfo  [0]{\@secondoftwo}%
\providecommand \bibfield  [0]{\@secondoftwo}%
\providecommand \translation [1]{[#1]}%
\providecommand \BibitemOpen [0]{}%
\providecommand \bibitemStop [0]{}%
\providecommand \bibitemNoStop [0]{.\EOS\space}%
\providecommand \EOS [0]{\spacefactor3000\relax}%
\providecommand \BibitemShut  [1]{\csname bibitem#1\endcsname}%
\let\auto@bib@innerbib\@empty
\bibitem [{\citenamefont {Pearl}(2000)}]{Pearl_book}%
  \BibitemOpen
  \bibfield  {author} {\bibinfo {author} {\bibfnamefont {J.}~\bibnamefont
  {Pearl}},\ }\href@noop {} {\emph {\bibinfo {title} {Causality: models,
  reasoning and inference}}}\ (\bibinfo  {publisher} {Cambridge Univ. Press},\
  \bibinfo {address} {New York},\ \bibinfo {year} {2000})\BibitemShut {NoStop}%
\bibitem [{\citenamefont {Spirtes}\ \emph {et~al.}(2000)\citenamefont
  {Spirtes}, \citenamefont {Glymour},\ and\ \citenamefont
  {Scheines}}]{Spirtes_book}%
  \BibitemOpen
  \bibfield  {author} {\bibinfo {author} {\bibfnamefont {P.}~\bibnamefont
  {Spirtes}}, \bibinfo {author} {\bibfnamefont {C.}~\bibnamefont {Glymour}}, \
  and\ \bibinfo {author} {\bibfnamefont {R.}~\bibnamefont {Scheines}},\
  }\href@noop {} {\emph {\bibinfo {title} {Causation, prediction, and
  search}}}\ (\bibinfo  {publisher} {MIT Press},\ \bibinfo {address}
  {Cambridge},\ \bibinfo {year} {2000})\BibitemShut {NoStop}%
\bibitem [{\citenamefont {Fritz}(2012)}]{Fritz_2012}%
  \BibitemOpen
  \bibfield  {author} {\bibinfo {author} {\bibfnamefont {T.}~\bibnamefont
  {Fritz}},\ }\href {arXiv:1206.5115} {\bibfield  {journal} {\bibinfo
  {journal} {New J. Phys.}\ }\textbf {\bibinfo {volume} {14}},\ \bibinfo
  {pages} {103001} (\bibinfo {year} {2012})}\BibitemShut {NoStop}%
\bibitem [{\citenamefont {Wood}\ and\ \citenamefont
  {Spekkens}(2015)}]{WoodSpekkens_2015}%
  \BibitemOpen
  \bibfield  {author} {\bibinfo {author} {\bibfnamefont {C.~J.}\ \bibnamefont
  {Wood}}\ and\ \bibinfo {author} {\bibfnamefont {R.~W.}\ \bibnamefont
  {Spekkens}},\ }\href@noop {} {\bibfield  {journal} {\bibinfo  {journal} {New
  Journal of Physics}\ }\textbf {\bibinfo {volume} {17}},\ \bibinfo {pages}
  {033002} (\bibinfo {year} {2015})}\BibitemShut {NoStop}%
\bibitem [{\citenamefont {Chaves}\ \emph {et~al.}(2014)\citenamefont {Chaves},
  \citenamefont {Luft},\ and\ \citenamefont {Gross}}]{ChavesEtAl_2014}%
  \BibitemOpen
  \bibfield  {author} {\bibinfo {author} {\bibfnamefont {R.}~\bibnamefont
  {Chaves}}, \bibinfo {author} {\bibfnamefont {L.}~\bibnamefont {Luft}}, \ and\
  \bibinfo {author} {\bibfnamefont {D.}~\bibnamefont {Gross}},\ }\href@noop {}
  {\bibfield  {journal} {\bibinfo  {journal} {New Journal of Physics}\ }\textbf
  {\bibinfo {volume} {16}},\ \bibinfo {pages} {043001} (\bibinfo {year}
  {2014})}\BibitemShut {NoStop}%
\bibitem [{\citenamefont {Chaves}\ \emph {et~al.}(2015)\citenamefont {Chaves},
  \citenamefont {Majenz},\ and\ \citenamefont {Gross}}]{ChavesEtAl_2015}%
  \BibitemOpen
  \bibfield  {author} {\bibinfo {author} {\bibfnamefont {R.}~\bibnamefont
  {Chaves}}, \bibinfo {author} {\bibfnamefont {C.}~\bibnamefont {Majenz}}, \
  and\ \bibinfo {author} {\bibfnamefont {D.}~\bibnamefont {Gross}},\
  }\href@noop {} {\bibfield  {journal} {\bibinfo  {journal} {Nature
  Communications}\ }\textbf {\bibinfo {volume} {6}},\ \bibinfo {pages} {5766}
  (\bibinfo {year} {2015})}\BibitemShut {NoStop}%
\bibitem [{\citenamefont {Markopoulou}\ and\ \citenamefont
  {Smolin}(1998)}]{MarkopoulouSmolin_1998}%
  \BibitemOpen
  \bibfield  {author} {\bibinfo {author} {\bibfnamefont {F.}~\bibnamefont
  {Markopoulou}}\ and\ \bibinfo {author} {\bibfnamefont {L.}~\bibnamefont
  {Smolin}},\ }\href {\doibase 10.1103/PhysRevD.58.084032} {\bibfield
  {journal} {\bibinfo  {journal} {Phys. Rev. D}\ }\textbf {\bibinfo {volume}
  {58}},\ \bibinfo {pages} {084032} (\bibinfo {year} {1998})}\BibitemShut
  {NoStop}%
\bibitem [{\citenamefont {Hardy}(2007{\natexlab{a}})}]{Hardy_2007b}%
  \BibitemOpen
  \bibfield  {author} {\bibinfo {author} {\bibfnamefont {L.}~\bibnamefont
  {Hardy}},\ }\href {http://stacks.iop.org/1751-8121/40/i=12/a=S12} {\bibfield
  {journal} {\bibinfo  {journal} {Journal of Physics A: Mathematical and
  Theoretical}\ }\textbf {\bibinfo {volume} {40}},\ \bibinfo {pages} {3081}
  (\bibinfo {year} {2007}{\natexlab{a}})}\BibitemShut {NoStop}%
\bibitem [{\citenamefont {Ried}\ \emph {et~al.}(2015)\citenamefont {Ried},
  \citenamefont {Agnew}, \citenamefont {Vermeyden}, \citenamefont {Janzing},
  \citenamefont {Spekkens},\ and\ \citenamefont {Resch}}]{RiedEtAl_2015}%
  \BibitemOpen
  \bibfield  {author} {\bibinfo {author} {\bibfnamefont {K.}~\bibnamefont
  {Ried}}, \bibinfo {author} {\bibfnamefont {M.}~\bibnamefont {Agnew}},
  \bibinfo {author} {\bibfnamefont {L.}~\bibnamefont {Vermeyden}}, \bibinfo
  {author} {\bibfnamefont {D.}~\bibnamefont {Janzing}}, \bibinfo {author}
  {\bibfnamefont {R.~W.}\ \bibnamefont {Spekkens}}, \ and\ \bibinfo {author}
  {\bibfnamefont {K.~J.}\ \bibnamefont {Resch}},\ }\href@noop {} {\bibfield
  {journal} {\bibinfo  {journal} {Nat. Phys.}\ }\textbf {\bibinfo {volume}
  {11}},\ \bibinfo {pages} {414} (\bibinfo {year} {2015})}\BibitemShut
  {NoStop}%
\bibitem [{\citenamefont {Leifer}(2006)}]{Leifer_2006}%
  \BibitemOpen
  \bibfield  {author} {\bibinfo {author} {\bibfnamefont {M.~S.}\ \bibnamefont
  {Leifer}},\ }\href@noop {} {\bibfield  {journal} {\bibinfo  {journal} {Phys.
  Rev. A}\ }\textbf {\bibinfo {volume} {74}},\ \bibinfo {pages} {042310}
  (\bibinfo {year} {2006})}\BibitemShut {NoStop}%
\bibitem [{\citenamefont {Aharonov}\ \emph {et~al.}(2009)\citenamefont
  {Aharonov}, \citenamefont {Popescu}, \citenamefont {Tollaksen},\ and\
  \citenamefont {Vaidman}}]{AharonovEtAl_2009}%
  \BibitemOpen
  \bibfield  {author} {\bibinfo {author} {\bibfnamefont {Y.}~\bibnamefont
  {Aharonov}}, \bibinfo {author} {\bibfnamefont {S.}~\bibnamefont {Popescu}},
  \bibinfo {author} {\bibfnamefont {J.}~\bibnamefont {Tollaksen}}, \ and\
  \bibinfo {author} {\bibfnamefont {L.}~\bibnamefont {Vaidman}},\ }\href@noop
  {} {\bibfield  {journal} {\bibinfo  {journal} {Phys. Rev. A}\ }\textbf
  {\bibinfo {volume} {79}},\ \bibinfo {pages} {052110} (\bibinfo {year}
  {2009})}\BibitemShut {NoStop}%
\bibitem [{\citenamefont {Chiribella}\ \emph {et~al.}(2009)\citenamefont
  {Chiribella}, \citenamefont {D'Ariano},\ and\ \citenamefont
  {Perinotti}}]{ChiribellaEtAl_2009}%
  \BibitemOpen
  \bibfield  {author} {\bibinfo {author} {\bibfnamefont {G.}~\bibnamefont
  {Chiribella}}, \bibinfo {author} {\bibfnamefont {G.~M.}\ \bibnamefont
  {D'Ariano}}, \ and\ \bibinfo {author} {\bibfnamefont {P.}~\bibnamefont
  {Perinotti}},\ }\href@noop {} {\bibfield  {journal} {\bibinfo  {journal}
  {Phys. Rev. A}\ }\textbf {\bibinfo {volume} {80}},\ \bibinfo {pages} {022339}
  (\bibinfo {year} {2009})}\BibitemShut {NoStop}%
\bibitem [{\citenamefont {Hardy}(2012)}]{Hardy_2012}%
  \BibitemOpen
  \bibfield  {author} {\bibinfo {author} {\bibfnamefont {L.}~\bibnamefont
  {Hardy}},\ }\href@noop {} {\bibfield  {journal} {\bibinfo  {journal} {Philos.
  T. Roy. Soc. A}\ }\textbf {\bibinfo {volume} {370}},\ \bibinfo {pages} {3385}
  (\bibinfo {year} {2012})}\BibitemShut {NoStop}%
\bibitem [{\citenamefont {Oreshkov}\ \emph {et~al.}(2012)\citenamefont
  {Oreshkov}, \citenamefont {Costa},\ and\ \citenamefont
  {Brukner}}]{Oreshkov2012}%
  \BibitemOpen
  \bibfield  {author} {\bibinfo {author} {\bibfnamefont {O.}~\bibnamefont
  {Oreshkov}}, \bibinfo {author} {\bibfnamefont {F.}~\bibnamefont {Costa}}, \
  and\ \bibinfo {author} {\bibfnamefont {C.}~\bibnamefont {Brukner}},\
  }\href@noop {} {\bibfield  {journal} {\bibinfo  {journal} {Nat. Commun.}\
  }\textbf {\bibinfo {volume} {3}},\ \bibinfo {pages} {1092} (\bibinfo {year}
  {2012})}\BibitemShut {NoStop}%
\bibitem [{\citenamefont {Fitzsimons}\ \emph {et~al.}(2013)\citenamefont
  {Fitzsimons}, \citenamefont {Jones},\ and\ \citenamefont
  {Vedral}}]{Fitzsimons2013}%
  \BibitemOpen
  \bibfield  {author} {\bibinfo {author} {\bibfnamefont {J.}~\bibnamefont
  {Fitzsimons}}, \bibinfo {author} {\bibfnamefont {J.}~\bibnamefont {Jones}}, \
  and\ \bibinfo {author} {\bibfnamefont {V.}~\bibnamefont {Vedral}},\
  }\href@noop {} {\bibfield  {journal} {\bibinfo  {journal} {arXiv:1302.2731}\
  } (\bibinfo {year} {2013})}\BibitemShut {NoStop}%
\bibitem [{\citenamefont {Leifer}\ and\ \citenamefont
  {Spekkens}(2013)}]{LeiferSpekkens_2013}%
  \BibitemOpen
  \bibfield  {author} {\bibinfo {author} {\bibfnamefont {M.}~\bibnamefont
  {Leifer}}\ and\ \bibinfo {author} {\bibfnamefont {R.~W.}\ \bibnamefont
  {Spekkens}},\ }\href@noop {} {\bibfield  {journal} {\bibinfo  {journal}
  {Phys. Rev. A}\ }\textbf {\bibinfo {volume} {88}},\ \bibinfo {pages} {052130}
  (\bibinfo {year} {2013})}\BibitemShut {NoStop}%
\bibitem [{\citenamefont {Chiribella}(2012)}]{Chiribella_2012}%
  \BibitemOpen
  \bibfield  {author} {\bibinfo {author} {\bibfnamefont {G.}~\bibnamefont
  {Chiribella}},\ }\href {\doibase 10.1103/PhysRevA.86.040301} {\bibfield
  {journal} {\bibinfo  {journal} {Phys. Rev. A}\ }\textbf {\bibinfo {volume}
  {86}},\ \bibinfo {pages} {040301} (\bibinfo {year} {2012})}\BibitemShut
  {NoStop}%
\bibitem [{\citenamefont {Hardy}(2007{\natexlab{b}})}]{Hardy_2007}%
  \BibitemOpen
  \bibfield  {author} {\bibinfo {author} {\bibfnamefont {L.}~\bibnamefont
  {Hardy}},\ }\href {arXiv:quant-ph/0701019} {\enquote {\bibinfo {title}
  {Quantum gravity computers: On the theory of computation with indefinite
  causal structure},}\ } (\bibinfo {year} {2007}{\natexlab{b}})\BibitemShut
  {NoStop}%
\bibitem [{\citenamefont {Ara\'ujo}\ \emph {et~al.}(2014)\citenamefont
  {Ara\'ujo}, \citenamefont {Costa},\ and\ \citenamefont
  {Brukner}}]{AraujoEtAl_2014}%
  \BibitemOpen
  \bibfield  {author} {\bibinfo {author} {\bibfnamefont {M.}~\bibnamefont
  {Ara\'ujo}}, \bibinfo {author} {\bibfnamefont {F.}~\bibnamefont {Costa}}, \
  and\ \bibinfo {author} {\bibfnamefont {v.}~\bibnamefont {Brukner}},\ }\href
  {\doibase 10.1103/PhysRevLett.113.250402} {\bibfield  {journal} {\bibinfo
  {journal} {Phys. Rev. Lett.}\ }\textbf {\bibinfo {volume} {113}},\ \bibinfo
  {pages} {250402} (\bibinfo {year} {2014})}\BibitemShut {NoStop}%
\bibitem [{\citenamefont {Procopio}\ \emph {et~al.}(2015)\citenamefont
  {Procopio}, \citenamefont {Moqanaki}, \citenamefont {Ara�jo}, \citenamefont
  {Costa}, \citenamefont {Calafell}, \citenamefont {Dowd}, \citenamefont
  {Hamel}, \citenamefont {Rozema}, \citenamefont {\u{C}aslav Brukner},\ and\
  \citenamefont {Walther}}]{ProcopioEtAl_2015}%
  \BibitemOpen
  \bibfield  {author} {\bibinfo {author} {\bibfnamefont {L.~M.}\ \bibnamefont
  {Procopio}}, \bibinfo {author} {\bibfnamefont {A.}~\bibnamefont {Moqanaki}},
  \bibinfo {author} {\bibfnamefont {M.}~\bibnamefont {Ara�jo}}, \bibinfo
  {author} {\bibfnamefont {F.}~\bibnamefont {Costa}}, \bibinfo {author}
  {\bibfnamefont {I.~A.}\ \bibnamefont {Calafell}}, \bibinfo {author}
  {\bibfnamefont {E.~G.}\ \bibnamefont {Dowd}}, \bibinfo {author}
  {\bibfnamefont {D.~R.}\ \bibnamefont {Hamel}}, \bibinfo {author}
  {\bibfnamefont {L.~A.}\ \bibnamefont {Rozema}}, \bibinfo {author}
  {\bibnamefont {\u{C}aslav Brukner}}, \ and\ \bibinfo {author} {\bibfnamefont
  {P.}~\bibnamefont {Walther}},\ }\href@noop {} {\bibfield  {journal} {\bibinfo
   {journal} {Nat. Commun.}\ }\textbf {\bibinfo {volume} {6}},\ \bibinfo
  {pages} {7913} (\bibinfo {year} {2015})}\BibitemShut {NoStop}%
\bibitem [{\citenamefont {Choi}(1975)}]{Choi1975}%
  \BibitemOpen
  \bibfield  {author} {\bibinfo {author} {\bibfnamefont {M.~D.}\ \bibnamefont
  {Choi}},\ }\href@noop {} {\bibfield  {journal} {\bibinfo  {journal} {Linear
  Algebra Appl.}\ }\textbf {\bibinfo {volume} {10}},\ \bibinfo {pages} {285}
  (\bibinfo {year} {1975})}\BibitemShut {NoStop}%
\bibitem [{\citenamefont {\u{C}ernoch}\ \emph {et~al.}(2008)\citenamefont
  {\u{C}ernoch}, \citenamefont {Soubusta}, \citenamefont
  {Bart\r{u}\v{s}kov\'a}, \citenamefont {Du\v{s}ek},\ and\ \citenamefont
  {Fiur\'a\v{s}ek}}]{CernochEtAl_2008}%
  \BibitemOpen
  \bibfield  {author} {\bibinfo {author} {\bibfnamefont {A.}~\bibnamefont
  {\u{C}ernoch}}, \bibinfo {author} {\bibfnamefont {J.}~\bibnamefont
  {Soubusta}}, \bibinfo {author} {\bibfnamefont {L.}~\bibnamefont
  {Bart\r{u}\v{s}kov\'a}}, \bibinfo {author} {\bibfnamefont {M.}~\bibnamefont
  {Du\v{s}ek}}, \ and\ \bibinfo {author} {\bibfnamefont {J.}~\bibnamefont
  {Fiur\'a\v{s}ek}},\ }\href {\doibase 10.1103/PhysRevLett.100.180501}
  {\bibfield  {journal} {\bibinfo  {journal} {Phys. Rev. Lett.}\ }\textbf
  {\bibinfo {volume} {100}},\ \bibinfo {pages} {180501} (\bibinfo {year}
  {2008})}\BibitemShut {NoStop}%
\bibitem [{\citenamefont {Jozsa}(1994)}]{jozsa_fidelity_1994}%
  \BibitemOpen
  \bibfield  {author} {\bibinfo {author} {\bibfnamefont {R.}~\bibnamefont
  {Jozsa}},\ }\href {\doibase 10.1080/09500349414552171} {\bibfield  {journal}
  {\bibinfo  {journal} {Journal of Modern Optics}\ }\textbf {\bibinfo {volume}
  {41}},\ \bibinfo {pages} {2315} (\bibinfo {year} {1994})}\BibitemShut
  {NoStop}%
\bibitem [{\citenamefont {Vidal}\ and\ \citenamefont
  {Werner}(2002)}]{VidalWerner_2002}%
  \BibitemOpen
  \bibfield  {author} {\bibinfo {author} {\bibfnamefont {G.}~\bibnamefont
  {Vidal}}\ and\ \bibinfo {author} {\bibfnamefont {R.~F.}\ \bibnamefont
  {Werner}},\ }\href {\doibase 10.1103/PhysRevA.65.032314} {\bibfield
  {journal} {\bibinfo  {journal} {Phys. Rev. A}\ }\textbf {\bibinfo {volume}
  {65}},\ \bibinfo {pages} {032314} (\bibinfo {year} {2002})}\BibitemShut
  {NoStop}%
\bibitem [{\citenamefont {Ara\'ujo}\ \emph {et~al.}(2015)\citenamefont
  {Ara\'ujo}, \citenamefont {Branciard}, \citenamefont {Costa}, \citenamefont
  {Feix}, \citenamefont {Giarmatzi},\ and\ \citenamefont {\v{C}aslav
  Brukner}}]{AraujoEtAl_2015}%
  \BibitemOpen
  \bibfield  {author} {\bibinfo {author} {\bibfnamefont {M.}~\bibnamefont
  {Ara\'ujo}}, \bibinfo {author} {\bibfnamefont {C.}~\bibnamefont {Branciard}},
  \bibinfo {author} {\bibfnamefont {F.}~\bibnamefont {Costa}}, \bibinfo
  {author} {\bibfnamefont {A.}~\bibnamefont {Feix}}, \bibinfo {author}
  {\bibfnamefont {C.}~\bibnamefont {Giarmatzi}}, \ and\ \bibinfo {author}
  {\bibnamefont {\v{C}aslav Brukner}},\ }\href
  {http://stacks.iop.org/1367-2630/17/i=10/a=102001} {\bibfield  {journal}
  {\bibinfo  {journal} {New Journal of Physics}\ }\textbf {\bibinfo {volume}
  {17}},\ \bibinfo {pages} {102001} (\bibinfo {year} {2015})}\BibitemShut
  {NoStop}%
\bibitem [{\citenamefont {Ángel Rivas}\ \emph {et~al.}(2014)\citenamefont
  {Ángel Rivas}, \citenamefont {Huelga},\ and\ \citenamefont
  {Plenio}}]{RivasEtAl_2014}%
  \BibitemOpen
  \bibfield  {author} {\bibinfo {author} {\bibnamefont {Ángel Rivas}}, \bibinfo
  {author} {\bibfnamefont {S.~F.}\ \bibnamefont {Huelga}}, \ and\ \bibinfo
  {author} {\bibfnamefont {M.~B.}\ \bibnamefont {Plenio}},\ }\href
  {http://stacks.iop.org/0034-4885/77/i=9/a=094001} {\bibfield  {journal}
  {\bibinfo  {journal} {Reports on Progress in Physics}\ }\textbf {\bibinfo
  {volume} {77}},\ \bibinfo {pages} {094001} (\bibinfo {year}
  {2014})}\BibitemShut {NoStop}%
\bibitem [{\citenamefont {Vacchini}\ \emph {et~al.}(2011)\citenamefont
  {Vacchini}, \citenamefont {Smirne}, \citenamefont {Laine}, \citenamefont
  {Piilo},\ and\ \citenamefont {Breuer}}]{VacchiniEtAl_2011}%
  \BibitemOpen
  \bibfield  {author} {\bibinfo {author} {\bibfnamefont {B.}~\bibnamefont
  {Vacchini}}, \bibinfo {author} {\bibfnamefont {A.}~\bibnamefont {Smirne}},
  \bibinfo {author} {\bibfnamefont {E.-M.}\ \bibnamefont {Laine}}, \bibinfo
  {author} {\bibfnamefont {J.}~\bibnamefont {Piilo}}, \ and\ \bibinfo {author}
  {\bibfnamefont {H.-P.}\ \bibnamefont {Breuer}},\ }\href@noop {} {\bibfield
  {journal} {\bibinfo  {journal} {New J. Phys.}\ }\textbf {\bibinfo {volume}
  {13}},\ \bibinfo {pages} {093004} (\bibinfo {year} {2011})}\BibitemShut
  {NoStop}%
\bibitem [{\citenamefont {Lavoie}\ \emph {et~al.}(2009)\citenamefont {Lavoie},
  \citenamefont {Kaltenbaek},\ and\ \citenamefont
  {Resch}}]{lavoie_experimental_2009}%
  \BibitemOpen
  \bibfield  {author} {\bibinfo {author} {\bibfnamefont {J.}~\bibnamefont
  {Lavoie}}, \bibinfo {author} {\bibfnamefont {R.}~\bibnamefont {Kaltenbaek}},
  \ and\ \bibinfo {author} {\bibfnamefont {K.~J.}\ \bibnamefont {Resch}},\
  }\href {\doibase 10.1088/1367-2630/11/7/073051} {\bibfield  {journal}
  {\bibinfo  {journal} {New Journal of Physics}\ }\textbf {\bibinfo {volume}
  {11}},\ \bibinfo {pages} {073051} (\bibinfo {year} {2009})}\BibitemShut
  {NoStop}%
\bibitem [{\citenamefont {Jamio{\l}kowski}(1972)}]{Jamiolkowski1972}%
  \BibitemOpen
  \bibfield  {author} {\bibinfo {author} {\bibfnamefont {A.}~\bibnamefont
  {Jamio{\l}kowski}},\ }\href@noop {} {\bibfield  {journal} {\bibinfo
  {journal} {Rep. Math. Phys.}\ }\textbf {\bibinfo {volume} {3}},\ \bibinfo
  {pages} {275} (\bibinfo {year} {1972})}\BibitemShut {NoStop}%
\bibitem [{\citenamefont {James}\ \emph {et~al.}(2001)\citenamefont {James},
  \citenamefont {Kwiat}, \citenamefont {Munro},\ and\ \citenamefont
  {White}}]{james_measurement_2001}%
  \BibitemOpen
  \bibfield  {author} {\bibinfo {author} {\bibfnamefont {D.~F.~V.}\
  \bibnamefont {James}}, \bibinfo {author} {\bibfnamefont {P.~G.}\ \bibnamefont
  {Kwiat}}, \bibinfo {author} {\bibfnamefont {W.~J.}\ \bibnamefont {Munro}}, \
  and\ \bibinfo {author} {\bibfnamefont {A.~G.}\ \bibnamefont {White}},\ }\href
  {\doibase 10.1103/PhysRevA.64.052312} {\bibfield  {journal} {\bibinfo
  {journal} {Physical Review A}\ }\textbf {\bibinfo {volume} {64}},\ \bibinfo
  {pages} {052312} (\bibinfo {year} {2001})}\BibitemShut {NoStop}%
\bibitem [{\citenamefont {Cover}\ and\ \citenamefont
  {Thomas}(2006)}]{CoverThomas_book}%
  \BibitemOpen
  \bibfield  {author} {\bibinfo {author} {\bibfnamefont {T.~M.}\ \bibnamefont
  {Cover}}\ and\ \bibinfo {author} {\bibfnamefont {J.~A.}\ \bibnamefont
  {Thomas}},\ }\href@noop {} {\emph {\bibinfo {title} {Elements of information
  theory}}}\ (\bibinfo  {publisher} {Wiley},\ \bibinfo {year}
  {2006})\BibitemShut {NoStop}%
\bibitem [{\citenamefont {Chiribella}\ \emph {et~al.}(2013)\citenamefont
  {Chiribella}, \citenamefont {D'Ariano}, \citenamefont {Perinotti},\ and\
  \citenamefont {Valiron}}]{ChiribellaEtAl_2013}%
  \BibitemOpen
  \bibfield  {author} {\bibinfo {author} {\bibfnamefont {G.}~\bibnamefont
  {Chiribella}}, \bibinfo {author} {\bibfnamefont {G.~M.}\ \bibnamefont
  {D'Ariano}}, \bibinfo {author} {\bibfnamefont {P.}~\bibnamefont {Perinotti}},
  \ and\ \bibinfo {author} {\bibfnamefont {B.}~\bibnamefont {Valiron}},\ }\href
  {\doibase 10.1103/PhysRevA.88.022318} {\bibfield  {journal} {\bibinfo
  {journal} {Phys. Rev. A}\ }\textbf {\bibinfo {volume} {88}},\ \bibinfo
  {pages} {022318} (\bibinfo {year} {2013})}\BibitemShut {NoStop}%
\end{thebibliography}%
\section*{Supplementary Notes}
\subsection*{Supplementary Note 1: Probabilistic mixtures of common-cause and cause-effect relations}\label{sec:ProbMixCCCE}

As defined in the main text of the article, a causal map $\mathcal{E}_{CB|D}$ is said to {physically realize} a {probabilistic mixture} of cause-effect and common-cause relations if it is possible to express it as follows: there is a hidden classical control variable, $J$, which only influences $B$, such that for every value of $J$,  either $B$ depends only on $D$ in the causal map or $B$ depends only on its common cause with $C$.  In this section, we discuss why this is the appropriate notion of probabilistic mixture to study.  We also demonstrate that it implies that the causal map has the form 
\begin{align}
\mathcal{E}_{CB|D} = w \mathcal{E} _{B|D}\otimes \rho_C +(1-w) \rho_{CB} \otimes {\rm Tr}_D,
\label{generalformprobmix}
\end{align} 
where $0 \le w\le 1$ and 
\begin{align}
{\rm Tr}_B \rho_{CB}=\rho_C.
\label{fixedmarg}
\end{align} 
 
 We are here concerned with what sorts of probabilistic mixtures of causal structures are physically realizable.   Note that a probabilistic mixture of alternatives is always physically realized by making the choice of alternatives depend causally on the value of a control variable that has been sampled from some probability distribution and for which the value is not observed.   It follows that to be physically realizable, a probabilistic mixture of the elements of a set of causal maps, $\{ \mathcal{E}^{(j)} _{CB|D}\}$, must have the form $\mathcal{E}_{CB|D} = \sum_{j} P(j) \mathcal{E}^{(j)} _{CB|D}$, where $J$ denotes the hidden control variable and $P(j)$ is the probability that $J=j$.
 
What is critical to recognize is that the causal dependence of systems on the control variable cannot be treated abstractly but must instead be considered as part of the causal structure.  One can then ask whether one can infer any constraints on the causal structure of the probabilistic mixture from the causal structure of the elements appearing in the mixture.  We argue that there is indeed a very natural constraint:
\begin{quote}
If all of the causal maps in a probabilistic mixture describe the same causal relation, then their mixture should describe this causal relation as well. 
\end{quote}

Note, first of all, that one particular implication of this constraint is that a causal map that is a probabilistic mixture of purely cause-effect maps should be purely cause-effect. We now demonstrate how this constraint implies a restriction on the sorts of probabilistic mixtures that can be physically realized. 

To begin, we consider the possibility that the set of probabilistic mixtures of causal structures that are physically realizable is the full set of such mixtures.  In this case, the causal maps corresponding to physically-realizable probabilistic mixtures of cause-effect and common-cause relations would be those that are a convex sum of causal maps each of which is purely cause-effect or purely common-cause, that is, those of the form
\begin{align}
\mathcal{E}_{CB|D} = \sum_{j\in \mathcal{J}_1
} P(j) \mathcal{E}^{(j)} _{B|D}\otimes \rho^{(j)}_C + \sum_{j\in \mathcal{J}_2
 } P(j) \rho^{(j)}_{CB} \otimes {\rm Tr}_D,
\label{generalformprobmix2}
\end{align} 
where the set of values of $J$ are partitioned into two subsets, denoted $\mathcal{J}_1$ and $\mathcal{J}_2$, and $P$ is a probability distribution thereon. 

\begin{figure}[h]
  \centering
  \includegraphics[scale=0.6]{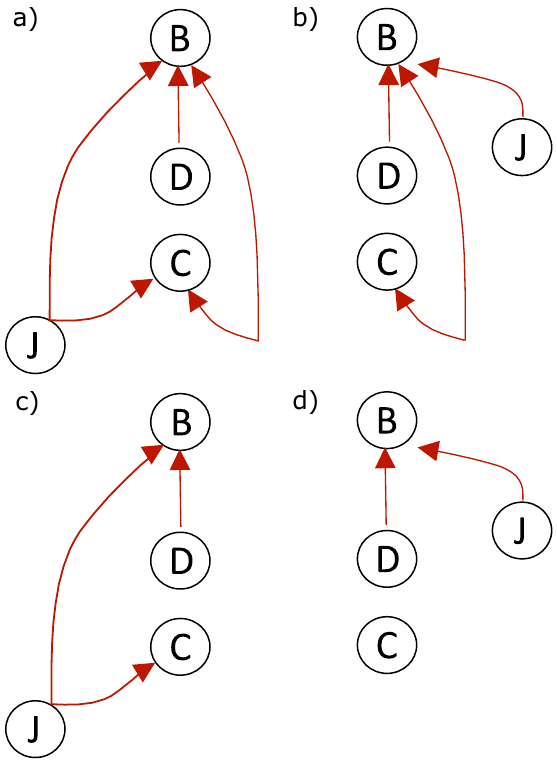}
  \caption{\footnotesize{\textbf{a)} Causal relations between $A$ (split into $C$ and
  $D$) and $B$ where the control variable $J$ acts as a common cause for both $B$ and
  $C$. \textbf{b)} We propose that a causal map represents a probabilistic mixture of common-cause
  and cause-effect if it is possible to achieve it by a hidden control variable $J$
  acting {only} on $B$. \textbf{c,d)} If one mixes only cause-effect relations, then the result, according to our constraint, should also be purely cause-effect. Again, this condition is not satisfied if the control variable $J$ acts as a common cause of $B$ and $C$, but it is if $J$ influences only $B$.}}
  \label{fig:controlvariableJ}
\end{figure}

In order to realize such a causal map, the control variable $J$ needs to have
a causal influence on both $B$ and $C$.  Otherwise, we could not explain how the
marginal states on $B$ and on $C$ both vary with the value of $J$.  In this case, $J$
acts as a common cause of $B$ and $C$, and the causal structure of the overall causal
map is that of \ref{fig:controlvariableJ}(a).  

The alternative proposal, the one that we endorse here, is that the control variable
$J$ only has a causal influence on $B$.  In this case, the causal structure of the
overall causal map is that of \ref{fig:controlvariableJ}(b).

We now demonstrate that the assumption that $J$ is a cause of both $B$ and $C$
violates the natural constraint articulated above.  Consider first the implication of
the constraint for probabilistic mixtures of purely cause-effect maps.  If every map
in the probabilistic mixture  is purely cause-effect, then for all values $j$ of $J$,
$\mathcal{E}^{(j)}_{CB|D} = \mathcal{E}^{(j)} _{B|D}\otimes \rho^{(j)}_C$.   The fact
that $J$ can influence $B$ and $C$ is encoded here in the fact that the marginal on
$C$ is $j$-dependent.  But now consider the causal map associated to this
probabilistic mixture.  It is $\mathcal{E}_{CB|D} = \sum_{ j} w_j \mathcal{E}^{(j)}
_{B|D}\otimes \rho^{(j)}_C$.  This is not a purely cause-effect map in general because, by
definition, such maps must take the form of a tensor product of a map from $D$ to $B$
and a state on $C$.  We can understand this by noting that the control variable acts
as a common cause, so that the causal structure is that depicted in \ref{fig:controlvariableJ}(c), which is clearly not purely cause-effect.  
Thus if one demands that any physical realization of a probabilistic mixture of purely cause-effect maps should itself be purely cause-effect, the scheme just described does not in fact realize such a mixture.  

On the other hand, if we demand that $J$ can only influence $B$, as in \ref{fig:controlvariableJ}(d), then $\mathcal{E}_{CB|D} = \left( \sum_{ j} \mathcal{E}^{(j)} _{B|D}\right) \otimes \rho_C$.  Here the marginal on $C$ is $j$-independent and therefore can be factorized out of the sum.  This {is} a purely cause-effect map, and so the constraint is satisfied.

We now show how to prove that all physically-realizable probabilistic mixtures can be expressed as a probabilistic mixture of just two causal maps, one of which is purely cause-effect and the other of which is purely common-cause. 

By our definition, it must be possible to partition the values of $J$ into two subsets, denoted $\mathcal{J}_1$ and $\mathcal{J}_2$, where for $j \in \mathcal{J}_1$, $B$ depends only on $D$, so that $\mathcal{E}^{(j)}_{CB|D} = \mathcal{E}^{(j)} _{B|D}\otimes \rho_C$,
and where for $j \in \mathcal{J}_2$, $B$ depends only on the common cause with $C$, so that $\mathcal{E}^{(j)}_{CB|D} = \rho^{(j)}_{CB} \otimes {\rm Tr}_D$.
The fact that the control variable $J$ is assumed to have no influence on $C$ implies that for all values of $J$, the causal map $\mathcal{E}^{(j)}_{CB|D}$ must have the same marginal on $C$.  This is why $\rho_C$  has no dependence  on $j$ in the expression for $\mathcal{E}^{(j)}_{CB|D}$ when $j \in \mathcal{J}_1$.  The lack of influence of $J$ on $C$ also implies that we must have 
\begin{equation}
{\rm Tr}_B \rho^{(j)}_{CB} = \rho_C,
\label{marg1}
\end{equation}
 for all $j\in \mathcal{J}_2$.

The overall causal map is obtained by weighting the $\mathcal{E}^{(j)}_{CD|D}$ by the probability $P(j)$ of their occurrence, so that 
$$\mathcal{E}_{CB|D}= \sum_{j \in \mathcal{J}_1} P(j) \mathcal{E}^{(j)}_{B|D} \otimes \rho_C +  \sum_{j \in \mathcal{J}_2} P(j) \rho^{(j)}_{CB} \otimes {\rm Tr}_D.$$
Finally, defining $$w \equiv \sum_{j \in \mathcal{J}_1} P(j),$$ and
$$\mathcal{E}_{B|D} \equiv \frac{1}{w} \sum_{j \in \mathcal{J}_1} P(j) \mathcal{E}^{(j)}_{B|D},$$ and 
\begin{align}
\rho_{CB} \equiv  \frac{1}{1-w} \sum_{j \in \mathcal{J}_2} P(j) \rho^{(j)}_{CB},
\label{rhoCB}
\end{align}
 we obtain supplementary equation~\ref{generalformprobmix}.  Supplementary
 equation~\ref{rhoCB} together with supplementary equation~\ref{marg1} implies
 supplementary equation~\ref{fixedmarg}.

\subsection*{Supplementary Note 2: The Choi isomorphism and different classes of causal maps}
\begin{figure}[h]
  \centering
  \includegraphics[scale=0.9]{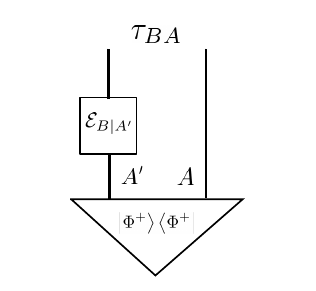}
  \caption{\footnotesize{Operational interpretation of the Choi state $\tau_{BA}$.
  It can be prepared by starting with the maximally entangled state $\ket{\Phi^+}$ on
  $A$ and $A'$ and applying the map to $A'$.}}
  \label{fig:choi-diagram}
\end{figure}

We begin by introducing a useful tool for defining and characterizing causal maps
that puts quantum channels, viz completely positive and trace-preserving (CPTP) maps,
on an equal footing with bipartite quantum states.
The Choi isomorphism \cite{Choi1975} (see also \cite{Jamiolkowski1972}) establishes
that completely positive maps from linear operators on the Hilbert space of $A$ to linear operators on $B$, denoted $\mathcal{E}_{B|A}:\mathcal{L}(\mathcal{H}_A)\rightarrow\mathcal{L}(\mathcal{H}_B)$, are isomorphic to positive-semidefinite operators $\tau_{BA}\in\mathcal{H}_B \otimes
\mathcal{H}_A$ given by
\begin{align}
\tau_{BA} \equiv \left(\mathcal{E}_{B|A'}\otimes \mathbb{1}_A \right) \left( \ket{\Phi^+}\bra{\Phi^+}_{A'A} \right).
\label{eq:Choi equation}
\end{align}
Here, $\ket{\Phi^+}_{A'A}=\tfrac{1}{\sqrt{d}}\sum_{k=1}^d\ket{k}_{A'}\ket{k}_A$ denotes the symmetric,
maximally entangled state between $A$ and an ancilla, $A'$, where $d$ is the Hilbert
space dimension of $A$ and $A'$. (Different choices of $\{\ket{k}\}$ lead to different forms of the Choi state; we will fix a convention for our calculations below.) Note that the different subscript in
$\mathcal{E}_{B|A'}$ indicates that the map is acting on input $A'$ and taking it to
 output $B$, as shown in \ref{fig:choi-diagram}.  Since $\ket{\Phi^+}$ is
normalized, $\tau_{BA}$ also has unit trace, making it a valid quantum state, if the map is trace-preserving. We
refer to $\tau_{BA}$ as the Choi state of the map $\mathcal{E}_{B|A}$.

The isomorphism also allows us to express the effect of the map on a generic input in terms of its Choi state~\cite{LeiferSpekkens_2013}: for any linear operator $\rho_A$ on $\mathcal{H}_A$,
\begin{align}
  \label{eq:Choi basic}
  \begin{split}
\mathcal{E}_{B|A}\left(\rho_A\right)
&=d~{\rm Tr}_A \left[ \left(T_A\tau_{BA}\right)\id_B\otimes \rho_A\right]\\ 
&=d ~{\rm Tr}_A \left[ \tau_{BA} \cdot \left( \id_B \otimes T_A\rho_A \right)\right].
\end{split}
\end{align}
The  transposition on $A$, denoted $T_A$, must be included in this expression in order for $\tau_{BA}$ to be a positive operator. It can be applied either to the input $\rho_A$ or to the Choi state itself. The identity operator on $B$, which is formally required in order for us to multiply $\tau_{BA}$ and $\rho_A$, is often omitted in the following for brevity.

\subsubsection*{Example classes of causal maps}

 In the following section, we show how the circuits presented in
 Fig.~\ref{fig:ExampleCircuits} realize examples of the causal maps
 $\mathcal{E}_{CB|D}$ in the classes \textsc{Coh} (hence \textsc{PhysQ}),
 \textsc{ProbQ}, \textsc{ProbC}, and \textsc{PhysC}, respectively.  In each case, we
 begin with a specification of the circuit elements, namely the state $\rho_{CE}$,
 which will be taken to be the maximally entangled state $\ket{\Phi^+}$
 in all following cases, and the gate $\mathcal{E}_{BF|DE}$, and find the causal map
 via 
 \begin{align}
\mathcal{E}_{CB|D} (\cdot) = {\rm Tr}_F \circ \mathcal{E}_{BF|DE} (\;\cdot \; \otimes \rho_{CE} ).
 \label{eq:general causal map}
 \end{align}
We then derive the associated Choi state,
 $\tau_{CBD}\in\mathcal{L(H}_C\otimes\mathcal{H}_B\otimes\mathcal{H}_D)$, which is given by
 \begin{align}
    \begin{split}
      \tau_{CBD}&=(\mathcal{E}_{CB|D'}\otimes\mathcal{I}_{D})\left(\ket{\Phi^+}\bra{\Phi^+}_{D'D}\right)\\
      &=\frac{1}{2}\sum_{j,k}\mathcal{E}_{CB|D'}\left(\ket{j}\bra{k}_{D'}\right)\otimes\ket{j}\bra{k}_D,
    \end{split}
    \label{eq:Choi-map}
  \end{align}
and use the witnesses introduced in the main paper to classify the corresponding causal maps.
We will take the basis $\ket{H}$, $\ket{V}$ of eigenstates of the Pauli operator $\sigma_z$ (which corresponds to horizontal and vertical polarization states) as the basis defining the Choi isomorphism; that is,
\begin{equation}
\ket{\Phi^+}  \equiv\frac{1}{\sqrt{2}}(\ket{HH}+\ket{VV}).
\end{equation}
  
\subsubsection*{Example of \textsc{Coh}}
 We begin with the circuit in Fig.~\ref{fig:ExampleCircuits}(d), which realizes an
 example of \textsc{Coh}. The circuit applies the partial swap gate from equation~\ref{eq:partialswap},
  \begin{align}
    \mathcal{E}_{BF|DE}(\cdot)=U_{BF|DE}~(\cdot)~U_{BF|DE}^\dag,
    \label{eq:partialswap1}
  \end{align}
  where
    \begin{align}
      U_{BF|DE}=
      \tfrac{1}{\sqrt{2}}\mathds{1}_{B|D} \otimes \mathds{1}_{F|E}
      + \tfrac{i}{\sqrt{2}}\mathds{1}_{B|E} \otimes \mathds{1}_{F|D}.
      \label{eq:partialswap2}
    \end{align}
 The first term of supplementary equation~\ref{eq:partialswap2} corresponds to the identity operation, which maps $D$ to $B$ and $E$ to $F$, and the second term corresponds to the swap operation, which maps $D$ to $F$ and $E$ to $B$.

    Inserting supplementary equation~\ref{eq:partialswap1} and supplementary equation~\ref{eq:partialswap2} as well as
    equation~\ref{eq:phiplus} from the main text into supplementary
    equation~\ref{eq:general causal map}, one can find an explicit expression for the
    causal map realized by the partial
    swap, $\mathcal{E}^{coh}_{CB|D}$. However, in order to derive the compact
    expression quoted in the main text (equation~\ref{eq:Ecoh}), we will consider a Kraus
    representation of the map: a set of operators $W_k$ such that 
  \begin{align}
  \mathcal{E}_{CB|D}^{coh}(\cdot)&=\sum_k W_k(\cdot)W_k^\dag.
  \end{align}
One possible choice of Kraus operators is given in terms of the unitary that defines $\mathcal{E}_{BF|DE}$ by
\begin{equation}
W_k = \bra{k_F} U_{BF|DE} \ket{\Phi^+}_{CE},
\end{equation}
where $\{\ket{k}\}$ is an arbitrary orthonormal basis of the Hilbert space of $F$. 

We take $k\in \{H,V\}$, ranging over eigenvectors of $\sigma_z$, which is the same basis that puts the initial state of $CE$ in a simple form. Substituting supplementary equation~\ref{eq:partialswap2}, we obtain
\begin{align}
W_k =& \sum_{m \in \{H,V\} } \bra{k_F} U_{BF|DE} \ket{mm}_{CE} \nonumber \\
=&  \sum_{m \in \{H,V\} } \tfrac{1}{\sqrt{2}} \bra{k_F} \mathds{1}_{B|D} \otimes \mathds{1}_{F|E} \ket{mm}_{CE}\nonumber\\
&+ \tfrac{i}{\sqrt{2}} \bra{k_F} \mathds{1}_{B|E} \otimes \mathds{1}_{F|D} \ket{mm}_{CE}\nonumber\\
=&\frac{1}{\sqrt{2}}(A_k+iB_k),
\end{align}
where we introduce the components
\begin{align}
  A_k &\equiv \sum_{m \in \{H,V\} }\tfrac{1}{\sqrt{2}} \bra{k_F} \mathds{1}_{B|D} \otimes \mathds{1}_{F|E} \ket{mm}_{CE} \nonumber \\
&= \tfrac{1}{\sqrt{2}}\mathds{1}_{B|D} \otimes \ket{k}_{C}\\
B_k &\equiv \sum_{m \in \{H,V\} } \tfrac{1}{\sqrt{2}}\bra{k_F} \mathds{1}_{B|E} \otimes \mathds{1}_{F|D} \ket{mm}_{CE} \nonumber\\ 
&= \ket{\Phi^+}_{CB} \bra{k}_D.
\end{align}
 Note that the $A_k$ contain only the two-qubit identity operator, which we expect to
 implement a purely cause-effect relation (the first term in
 supplementary equation~\ref{eq:partialswap2}), whereas the $B_k$ contain only the swap operator, which
 we expect to realize a purely common-cause relation (the second term in supplementary equation~\ref{eq:partialswap2}).

In terms of the $A_k$ and $B_k$, we have
  \begin{align}
	\mathcal{E}_{CB|D}^{coh}(\cdot)&=\sum_k\frac{1}{2}(A_k+iB_k)(\cdot)(A_k+iB_k)^\dag\\
	&=\sum_k
	\frac{1}{2}A_k(\cdot)A_k^\dag+\frac{1}{2}B_k(\cdot)B_k^\dag\nonumber\\
	\nonumber&\hspace{0.6cm}-\frac{i}{2}\left\{A_k(\cdot)B_k^\dag-B_k(\cdot)A_k^\dag\right\}.
      \label{}
    \end{align}
    The effect of the first term is 
    \begin{align}
    \sum_k A_k(\cdot)A_k^\dag &= \tfrac{1}{2}\sum_k \ket{k}\bra{k}_C \otimes \mathds{1}_{B|D} (\cdot) \mathds{1}^\dagger_{B|D} \nonumber \\
    &=\tfrac{1}{2}\mathds{1}_C \otimes \mathcal{I}_{B|D} (\cdot)\equiv \mathcal{E}^{ce}_{CB|D}(\cdot),
    \label{eq:Ece}
    \end{align}
    that is, it applies the identity channel from $D$ to $B$, which is our example of a purely cause-effect relation. 
    The effect of the second term is 
     \begin{align}
    \sum_k B_k(\cdot)B_k^\dag &= \ket{\Phi^+} \bra{\Phi^+}_{CB} \otimes \sum_k \bra{k_D}  (\cdot) \ket{k_D} \nonumber \\
    &=\ket{\Phi^+} \bra{\Phi^+}_{CB} {\rm Tr}_D (\cdot)\equiv \mathcal{E}^{cc}_{CB|D}(\cdot),
    \label{eq:Ecc}
    \end{align}
    that is, it traces out $D$ and prepares the state $\ket{\Phi^+}$ on $CB$, which is our example of a purely common-cause relation. 
    Finally, the cross terms take the form
    \begin{align}
      -\frac{i}{2}\sum_k &\left\{A_k(\cdot)B_k^\dag-B_k(\cdot)A_k^\dag\right\}\\
      \nonumber=&-\frac{i}{2}\sum_k\Big\{
      \left(\tfrac{1}{\sqrt{2}}\ket{k}_C\otimes\id_{B|D}\right)(\cdot)\left(\bra{\Phi^+}_{CB}\otimes\ket{k}_D\right)\\
      \nonumber&
      -\left(\ket{\Phi^+}_{CB}\otimes\bra{k}_D\right)(\cdot)\left(\tfrac{1}{\sqrt{2}}\bra{k}_C\otimes\id_{B|D}\right)\Big\}\\
      \nonumber =&-i \left[\tfrac{1}{2}\id_C \otimes
      \mathcal{I}_{B|D}(\cdot)\right]\ket{\Phi^+}\bra{\Phi^+}_{CB} \\
\nonumber & +i \ket{\Phi^+}\bra{\Phi^+}_{CB}\left[\tfrac{1}{2}\id_C\otimes
     \mathcal{I}_{B|D}(\cdot)\right]\Big\},
  \end{align}
  which completes the derivation of equation~\ref{eq:Ecoh} in the main paper. 
  
  The Choi representation of  this causal map is then calculated using supplementary equation~\ref{eq:Choi-map},
  \begin{align}
    \nonumber\tau^{coh}_{CBD}&=
    \frac{1}{2}\left(\tfrac{1}{2}\id_C\otimes\ket{\Phi^+}\bra{\Phi^+}_{BD}\right)
    +\frac{1}{2}\left(\ket{\Phi^+}\bra{\Phi^+}_{CB}\otimes\tfrac{1}{2}\id_D\right)\\
    \label{eq:taucoh}
    &-\frac{i}{2}\Big\{\left(\id_C\otimes\ket{\Phi^+}\bra{\Phi^+}_{BD}\right)\cdot\left(\ket{\Phi^+}\bra{\Phi^+}_{CB}\otimes\id_D\right)\\
   \nonumber
    &\hspace{0.5cm}
    -\left(\ket{\Phi^+}\bra{\Phi^+}_{CB}\otimes\id_D\right)\cdot\left(\id_C\otimes\ket{\Phi^+}\bra{\Phi^+}_{BD}\right)\Big\}.
  \end{align}

The Choi state in supplementary equation~\ref{eq:taucoh} contains all the information required to characterize the causal structure. In order to evaluate our witnesses, we will calculate the induced states associated with finding certain states on $C$, $B$ or $D$. Letting $\Pi^b$ denote the projector associated with an eigenvalue $b$ in a particular measurement on system $B$, we use
 \begin{align}
   \tau^b_{CD}=\frac{1}{{\rm Tr}\left[\Pi^b_B \cdot \tau^{coh}_{CBD} \right]}{\rm Tr}_B\left[\Pi^b_B \cdot \tau^{coh}_{CBD} \right],
 \end{align}
and similarly for a projector $\Pi^c$ on $C$,
 \begin{align}
   \tau^c_{BD}=\frac{1}{{\rm Tr}\left[\Pi^c_C \cdot \tau^{coh}_{CBD} \right]}{\rm Tr}_C\left[\Pi^c_C \cdot \tau^{coh}_{CBD} \right].
 \end{align}
The expression for the induced state on $CB$ generated by an input $\Pi^d$ on $D$
differs from the above by a partial transpose, and  the renormalization factor to ensure unit trace is always $d_D$:
\begin{align}
  \tau^d_{CB}=\mathcal{E}^{coh}_{CB|D}\left(\Pi^d_D\right)= d_D{\rm Tr}_D\left[\tau^{coh}_{CBD} T_D \left(\Pi^d_D\right) \right].
 \end{align}

If one prepares $D$ in the state $\ket{H}$, then the state on $CB$
 is $\tau^{H}_{CB} = \tfrac{3}{4} \ket{\psi}\bra{\psi} + \tfrac{1}{4} \ket{
 VH}\bra{VH}$ where $\ket{\psi} = \tfrac{2}{\sqrt{6}} \ket{HH} + e^{i\pi/4}
 \tfrac{1}{\sqrt{3}} \ket{VV} $, and one can see that $\tau^{H}_{CB}$ is entangled. The same holds if one prepares $\ket{V}$ on $D$ instead.
The causal map $\mathcal{E}_{CB|D}^{coh}$ is therefore quantum in the cause-effect pathway. If one measures $C$  and selects for the state
 $\ket{H}$, the resulting map from $D$ to $B$ is Choi-isomorphic (up to normalization) to the
 state $\tau^{H}_{BD} = \tfrac{3}{4} \ket{\tilde\psi}\bra{\tilde\psi}
 + \tfrac{1}{4}\ket{HV}\bra{HV}$, where $\ket{\tilde\psi}
 = \tfrac{2}{\sqrt{6}} \ket{HH} + e^{-i\pi/4} \tfrac{1}{\sqrt{3}} \ket{VV}$. One can see that $\tau^{H}_{BD}$ is entangled. The same result is found if the measurement on $C$ finds $\ket{V}$, and therefore the causal map is quantum in the common-cause pathway. 

Finding $B$ in the state $\ket{H}$ also induces entanglement between $C$ and $D$, as was already shown in the main text. For completeness, we note that, if one finds $\ket{V}$ instead, the induced Choi state on $CD$ is
 \begin{align}
   \tau_{CD}^{V}=&\frac{1}{2}
   \ket{VV}\bra{VV}+\frac{1}{2}\ket{\tilde\varphi}\bra{\tilde\varphi}, 
 \end{align} with
 $\ket{\tilde\varphi} \equiv  \tfrac{1}{\sqrt{2}} (\ket{HV}+i \ket{VH}$,
 which is also entangled. 
 As a result, the causal map $\mathcal{E}_{CB|D}^{coh}$ satisfies all the requirements for the class \textsc{Coh}: it is quantum in both the cause-effect and the common-cause pathways and, furthermore, exhibits a quantum Berkson effect.

 \subsubsection*{Example of \textsc{ProbQ}} 

 The circuit realizing an example of the class \textsc{ProbQ} is presented
 Fig.~\ref{fig:ExampleCircuits}(c). This circuit implements a probabilistic mixture of identity and swap,  
  \begin{align}
    \begin{split}
      \mathcal{E}_{BF|DE}(\cdot)&=\frac{1}{2}(\mathds{1}_{B|D} \otimes
      \mathds{1}_{F|E})~(\cdot)~\left(\mathds{1}_{B|D}
     \otimes \mathds{1}_{F|E}\right)\\
     &+\frac{1}{2}(\mathds{1}_{B|E} \otimes \mathds{1}_{F|D})~(\cdot)~\left(\mathds{1}_{B|E} \otimes
     \mathds{1}_{F|D}\right).
  \end{split}
  \label{eq:ProQ gate}
  \end{align}
  Using supplementary equation~\ref{eq:general causal map}, we find
    \begin{align}
      \mathcal{E}_{CB|D}(\cdot)=
      \frac{1}{2}\mathcal{E}^{ce}_{CB|D}(\cdot)
      +\frac{1}{2}\mathcal{E}^{cc}_{CB|D}(\cdot),
      \label{}
    \end{align}
    where $\mathcal{E}^{ce}_{CB|D}$ and $\mathcal{E}^{cc}_{CB|D}$ are defined in
    supplementary equation~\ref{eq:Ece} and supplementary equation~\ref{eq:Ecc} respectively. This shows explicitly that the map is a probabilistic mixture of a purely cause-effect term and a purely common-cause term.

  By supplementary equation~\ref{eq:Choi-map}, the Choi state is
  \begin{align}
    \tau_{CBD}&=
    \frac{1}{2}\tfrac{1}{2}\id_C\otimes\ket{\Phi^+}\bra{\Phi^+}_{BD}+\frac{1}{2}\ket{\Phi^+}\bra{\Phi^+}_{CB}\otimes\tfrac{1}{2}\id_D.
    \label{}
  \end{align}
  Finding $C$ in the  state $\ket{H}$ implies  $\tau_{BD}^{H}=\frac{1}{2}\ket{\Phi^+}\bra{\Phi^+}_{BD}+
  \frac{1}{2}\ket{H}\bra{H}_B\otimes\tfrac{1}{2}\id_D$, which is entangled, and similarly if $C$ is found in the state $\ket{V}$. The causal map is therefore quantum in the cause-effect pathway. If we prepare $\ket{H}$ on $D$, then the state on $CB$ is $\tau_{CB}^{H}=\frac{1}{2}\ket{\Phi^+}\bra{\Phi^+}_{CB}+ \frac{1}{2}\left(\tfrac{1}{2}\id_C\otimes\ket{H}\bra{H}_B\right)$, which is also entangled. The same holds when preparing $\ket{V}$ on $D$, and consequently  the causal map is quantum in the common-cause pathway. 
  It follows that the causal map is in the class \textsc{ProbQ}.

\subsubsection*{Causal maps that are classical on both pathways}

A general way to realize causal maps that are classical on both the cause-effect and common-cause pathway is to insert completely dephasing channels before and after the gate $\mathcal{E}_{BF|DE}$.
The generic completely dephasing channel takes the form
  \begin{align}
    \Delta_{\hat{n}}(\rho)=\frac{1}{2}\rho+\frac{1}{2}\left([\hat
    n\cdot \vec{\sigma}]\rho[\hat n\cdot \vec{\sigma}]\right), \label{eq:dephase}
  \end{align}
  where $\vec{\sigma}$ is the vector of Pauli observables and the Bloch vector $\hat{n}$ specifies the eigenbasis on which we dephase. The dephasing effectively reduces the qubits $B,D,E$, and $F$ to classical binary variables, which we denote  $b,d,e,f$, and reduces the map $\mathcal{E}_{BF|DE}$ to a conditional probability distribution $P(bf|de)$: letting $\ket{b}$ denote the elements of a preferred basis of $\mathcal{H}_B$ -- namely, the eigenbasis of $\hat n_B \cdot \vec{\sigma}$ --, and similarly for $F$, $D$ and $E$, we can write
  \begin{align}
  \begin{split}
    \mathcal{E}_{BF|DE}(\rho_{DE})= \sum_{b,d,e,f} & P(bf|de) \ket{b}\bra{b}\otimes{\ket{f}\bra{f}}\\
    &\times {\rm Tr}_{DE} \left( \ket{d}\bra{d}\otimes\ket{e}\bra{e}\rho_{DE}\right).
  \end{split}
  \label{eq:classical gate}
  \end{align}
  The dephasing on $E$ also effectively reduces $C$ to a classical binary variable, since $C$ is only related to other variables in the problem via $E$. We denote this variable by $c$ and the corresponding preferred basis (which generally depends on the initial joint state $\rho_{CE}$) by $\ket{c}$. 
   Substituting supplementary equation~\ref{eq:classical gate} into  supplementary equation~\ref{eq:general causal map}, we find that the causal map takes the form
  \begin{align}
  \mathcal{E}_{CB|D} (\rho_D) \equiv \sum_{c,b,d} P(cb|d)\ket{c}\bra{c}\otimes \ket{b}\bra{b}
  \times {\rm Tr}_{D} \left( \ket{d}\bra{d}\rho_D \right).
  \label{eq:classical map}
  \end{align}
  Note that, since the prescription for deriving the causal map $\mathcal{E}_{CB|D}$ from $ \mathcal{E}_{BF|DE}$ and $\rho_{CE}$ involves tracing out system $F$, we find this form of the causal map independently of whether we actually apply dephasing on $F$: dephasing on $D$, $E$ and $B$ is sufficient.
  
 It follows that the corresponding Choi state 
takes the form
 \begin{align}
      \tau_{CBD}=\sum_{c,b,d} P(cb|d) u(d)\ket{c}\bra{c}\otimes \ket{b}\bra{b}
  \otimes  \tilde{\ket{d}}\tilde{\bra{d}},
  \label{eq:classical choi}
    \end{align}
where $u(d)$ denotes the uniform distribution over $d$ and $\tilde{\ket{d}}$ is related to $\ket{d}$ by complex conjugation in the basis that defines the Choi isomorphism (in our case, $\{\ket{H},\ket{V}\}$). 
Similarly, the induced states, for a preparation 
$\rho_D=\ket{d}\bra{d}$, projection $\Pi^c_C=\ket{c}\bra{c}$, and projection $\Pi^b_B=\ket{b}\bra{b}$ are given by the operators
  \begin{align}
    \begin{split}
    &\tau^{d}_{CB} = \sum_{c,b} P^{d}(c,b)\ket{c}\bra{c}\otimes\ket{b}\bra{b}\\
    &\tau^{c}_{BD} = \sum_{b,d}
    P^{c}(b,d)\ket{b}\bra{b}\otimes\tilde{\ket{d}}\tilde{\bra{d}},\\
    &\tau^{b}_{CD} = \sum_{c,d}
    P^{b}(c,d)\ket{c}\bra{c}\otimes\tilde{\ket{d}}\tilde{\bra{d}}.
    \end{split}
  \label{eq:tausep}
  \end{align}
with 
  \begin{align}
    \begin{split}
    &P^{d}(c,b)=P(cb|d)\\
    &P^{c}(b,d)=P(cb|d)u(d)/\left[ \sum_{bd} P(cb|d)u(d)\right]\\
    &P^{b}(c,d)=P(cb|d)u(d)/\left[\sum_{cd} P(cb|d)u(d)\right].
    \end{split}
  \end{align}
Any operator of the form of supplementary equation~\ref{eq:tausep} is separable.  Therefore, by our criterion, such dephased causal maps are not quantum in either pathway.

   \subsubsection*{Example of \textsc{ProbC}}

A circuit realizing an example of \textsc{ProbC} is presented in Fig.~\ref{fig:ExampleCircuits}(a). It applies complete dephasing channels to $D$ and $E$ only. However, note that  $B$ and $F$ are obtained from $D$ and $E$ by either the two-qubit identity channel or the swap. This implies that $B$ and $F$ are also effectively classical, on the same bases on which we dephase $D$ and $E$.
 The gate therefore can be expressed in the form of supplementary equation~\ref{eq:classical gate}, with
   \begin{align}
     P(bf|de)=\frac{1}{2}\delta_{b,d}\delta_{f,e}+\frac{1}{2}\delta_{b,e}\delta_{f,d},
     \label{}
   \end{align}
 where $\delta_{x,y}$ denotes the Kronecker delta function over variables $x,y$. The variables $d$ and $e$ are mapped either to $b$ and $f$ or to $f$ and $b$, respectively, with equal probability: a probabilistic mixture of classical identity and swap. The causal map is therefore effectively described by a classical probability distribution, as in supplementary equation~\ref{eq:classical map}, with
 \begin{align}
   P(cb|d)=\frac{1}{2}u(c)\delta_{b,d}+\frac{1}{2}\delta_{c,b}u(c),
   \label{eq:ProbC}
 \end{align}
 where $u(x)$ denotes the uniform distribution of the variable $x$. The dephasing ensures that the common-cause and cause-effect components of the causal map are classical, and the form of $\mathcal{E}_{CB|D}$ makes it clear that this is a probabilistic mixture of classical cause-effect and common-cause relations. The causal map is therefore in the class \textsc{ProbC}.

 \subsubsection*{Example of \textsc{PhysC}}
 
 A circuit realizing an example of \textsc{PhysC} is presented in
 Fig.~\ref{fig:ExampleCircuits}(b). Again, we explicitly apply dephasing channels
 only to $D$ and $E$, but note that the classical XNOR gate, which generates $B$ in
 the left-hand panel, {implicitly} defines a preferred basis -- in other words:
 if $B$ is the output of a classical XNOR (Not-XOR: $b=-de$ for $d,e\in\{-1,1\}$),
 then $B$ must be (effectively) classical. In the right-hand panel, $B$ is prepared
 in the maximally
 mixed state, which can also be described as effectively classical. The same holds
 for $F$, and we can therefore again express the causal map in terms of a classical
 conditional distribution, 
 \begin{align}
    P(bf|de) =\frac{1}{2}\delta_{b,-de}u(f)+\frac{1}{2} u(b)\delta_{f,-de}~,
 \end{align}
The gate either sets $b=-de$ and generates $f$ at random or vice versa. This leads to
a causal map of the form of supplementary equation~\ref{eq:classical map}, with
  \begin{align}
  P(cb|d) =\frac{1}{2}u(c)u(b)+\frac{1}{2} u(c)\delta_{b,-cd}.
  \label{eq:PPhysC}
  \end{align}
Even though $b$ is completely unaffected by $d$ and $e$ in the first term, in the second term $b$ depends nontrivially on {both} inputs. One can see that this makes the causal map a physical mixture: indeed, the induced state $\tau^{b}_{CD}$ in this case is given by
 supplementary equation~\ref{eq:tausep} with
  \begin{align}
 P^{b}(cd) =\frac{1}{2}u(c)u(d)+\frac{1}{2} u(c) \delta_{b,-cd}.
  \end{align}
The mutual information between $c$ and $d$ in this distribution is 0.19 bits for
either value of $b$. By contrast, we will show in a later section that the induced
mutual information between binary variables $c$ and $d$ for any {probabilistic} mixture of common-cause and cause-effect with uniform
prior distributions (which is the case here) is at most  0.12 bits. It follows that the present example must be a physical mixture, and noting furthermore that the causal
 map is classical in both pathways, we conclude that it belongs to the class \textsc{PhysC}. 

\subsubsection*{Proof that \textsc{Coh} is a {strict} subset of \textsc{PhysQ}}

Based on the previous scenarios, one can now see that \textsc{Coh} is in fact a {strict} subset of \textsc{PhysQ}. To show this, we will explicitly construct a causal map that belongs to \textsc{PhysQ} but not to \textsc{Coh}. To wit, consider a probabilistic mixture of our examples of \textsc{ProbQ} and \textsc{PhysC}, with a small weight $\epsilon$ for the latter:
  \begin{align}
      \nonumber&\mathcal{E}_{CB|D}(\cdot)=\frac{1-\epsilon}{2}
      \Big(\tfrac{1}{2}\id_C\otimes\mathcal{I}_{B|D}(\cdot)
      +\ket{\Phi^+}\bra{\Phi^+}_{CB} \times{\rm Tr}_D(\cdot)\Big) \\
      & \hspace{1.cm}+ \frac{\epsilon}{2}\Big(\tfrac{1}{2}\id_C\otimes\tfrac{1}{2}\id_B \times{\rm Tr}_D(\cdot) \\ 
  &\nonumber \hspace{1.2cm}   +\sum_{c,b,d} u(c)\delta_{b,-cd}
  \ket{c}\bra{c}\otimes \ket{b}\bra{b} \times {\rm Tr}_{D} \left(
  \ket{d}\bra{d}~\cdot \right)\Big), 
    \end{align}
where we take the preferred bases for the example of \textsc{PhysC}, $\ket{c}$, $\ket{b}$ and $\ket{d}$, to each be the eigenbasis of $\sigma_z$.

In this case, the resulting causal map is still quantum on both pathways: indeed, finding $C$ in the
  state $\ket{H}$ implies
  \begin{align}
\tau_{BD}^{H}&=\frac{1-\epsilon}{2}\ket{\Phi^+}\bra{\Phi^+}_{BD}+
  \frac{1-\epsilon}{2}\ket{H}\bra{H}_B\otimes\tfrac{1}{2}\id_D \nonumber \\
& +  \frac{\epsilon}{2}\tfrac{1}{2}\id_B \otimes \tfrac{1}{2}\id_D 
+\frac{\epsilon}{2} \sum_{b,d} u(d) \delta_{b,-d} \ket{b}\bra{b}\otimes  \tilde{\ket{d}}\tilde{\bra{d}},
\end{align} 
which is entangled for a range of $\epsilon$, and similarly for finding $C$ in the state $\ket{V}$. The causal map is therefore quantum in the cause-effect
  pathway. If we prepare $\ket{H}$ on $D$, then the state on
  $CB$ is 
\begin{align}
\tau_{CB}^{H}&=\frac{1-\epsilon}{2}\ket{\Phi^+}\bra{\Phi^+}_{CB}+
  \frac{1-\epsilon}{2}\tfrac{1}{2}\id_C\otimes\ket{H}\bra{H}_B \nonumber  \\
&+  \frac{\epsilon}{2}\tfrac{1}{2}\id_C\otimes\tfrac{1}{2}\id_B   +\frac{\epsilon}{2} \sum_{c,b} u(c)\delta_{b,-c} \ket{c}\bra{c}\otimes \ket{b}\bra{b},
\end{align}
which is also entangled. The same holds for preparing $\ket{V}$ on $D$, and consequently  the causal map is quantum in the common-cause pathway. 
However, the causal map cannot be realized by a probabilistic mixture of purely common-cause and purely cause-effect relations, since the witness of physical mixture is $\epsilon/4$, i.e. non-zero for all valid values of $\epsilon$. (The fact that the last term, $b=-cd$, has $b$ depending simultaneously on $c$ and $d$ is also suggestive of a physical mixture, but not conclusive.)
Consequently, the map belongs to \textsc{PhysQ}. 

On the other hand, no measurement outcome on $B$ implies entanglement on $CD$: since every term in the expression for $\mathcal{E}_{CB|D}$ has the form of a tensor product between $C$ and $D$, the state induced when selecting for any state on $B$ must be separable. It follows that the map is not in \textsc{Coh}.

\subsection*{Supplementary Note 3: Implementing the examples using a single experimental set-up}

 In the following section, we show how the single set-up in Fig.~5 can experimentally realize our examples of the classes \textsc{Coh} (hence \textsc{PhysQ}), \textsc{ProbQ}, \textsc{ProbC}, and
 \textsc{PhysC}.  We first describe the implementation of the partial swap gate,
 which allows us to realize the example of \textsc{Coh} from
 Fig.~\ref{fig:ExampleCircuits}(d). Next we describe how this gate can be modified
 in order to realize the example of \textsc{ProbQ} from Fig.~\ref{fig:ExampleCircuits}(c). We then move on to discuss how applying complete dephasing channels on $D$, $E$ and $B$ yields causal maps where both pathways are classical.  We show that by modifying the type of dephasing, we can realize the examples of \textsc{ProbC} and
 \textsc{PhysC} (up to a sign change) presented in Fig.~\ref{fig:ExampleCircuits}(a,b). 

 \subsubsection*{Example of \textsc{Coh}}

 We begin by describing how the partial swap gate is implemented experimentally.  In
 the set-up of Fig.~5, two photons are input at $D$ and $E$ and
 measured in coincidence at $B$ and $F$. If the photons input at $D$ and $E$ are
 indistinguishable, then when they arrive at the first beam splitter, they will bunch
 if their polarization state lies in the symmetric subspace (spanned by the triplet
 basis), while they will anti-bunch if their polarization state lies in the
 anti-symmetric subspace (singlet). If they bunch, then a coincidence at $B$ and $F$
 can only be obtained if both photons take the clockwise path in the Sagnac
 interferometer.  However, if they anti-bunch, the two photons will take opposite
 paths. The photon travelling along counterclockwise path will acquire an extra
 phase, denoted $\theta$, due to the glass windows, while the other, on the clockwise
 path, acquires no extra phase.

 In this configuration, the gate applies a phase difference between the symmetric and anti-symmetric subspaces of the photon state. Recall that the projectors onto these subspaces can be written as linear combinations of the identity and swap operators,
 \begin{align}
\begin{split}
    \mathbb{S}=
    \frac{1}{2} \left(\mathds{1}_{B|D} \otimes \mathds{1}_{F|E}+\mathds{1}_{B|E} \otimes
    \mathds{1}_{F|D} \right)  \\
    \mathbb{A}=
    \frac{1}{2} \left(\mathds{1}_{B|D} \otimes \mathds{1}_{F|E}-\mathds{1}_{B|E} \otimes
    \mathds{1}_{F|D} \right),
    \label{eq:projectors}
\end{split}
  \end{align}
so that the gate takes the form
  \begin{align}
    \mathcal{E}_{BF|DE}(\rho)=\left(\mathbb{S}+e^{i\theta}\mathbb{A}\right)\rho
    \left(\mathbb{S}+e^{i\theta}\mathbb{A}\right)^\dag.
    \label{eq:cohgate}
  \end{align}
Substituting the expressions for $ \mathbb{S}$ and $ \mathbb{A}$, we find
  \begin{align}
    \mathbb{S}+&e^{i\theta}\mathbb{A}=\\
    =&\frac{\mathds{1}_{B|D} \otimes \mathds{1}_{F|E} +\mathds{1}_{B|E} \otimes    \mathds{1}_{F|D}}{2}  \nonumber \\
&+e^{i\theta}\frac{\mathds{1}_{B|D} \otimes
    \mathds{1}_{F|E}-\mathds{1}_{B|E} \otimes \mathds{1}_{F|D}}{2} \nonumber \\
    =&e^{i\theta/2}\left(\cos{(\theta/2)}\mathds{1}_{B|D} \otimes
    \mathds{1}_{F|E}-i\sin{(\theta/2)}\mathds{1}_{B|E} \otimes \mathds{1}_{F|D}\right) \nonumber 
  \end{align}

 Thus, by adjusting the phase of the Sagnac interferometer to $\theta=-\pi/2$, we
 obtain a gate $\mathcal{E}_{BF|DE}$ that implements the partial swap unitary
 $U_{BF|DE}$ given by equation \ref{eq:partialswap} in the main text. This allows us to
 build the circuit in Fig.~\ref{fig:ExampleCircuits}(d) and realize our example of the class \textsc{Coh}.

 \subsubsection*{Example of \textsc{ProbQ}}
 In order to realize our example of the class \textsc{ProbQ}, we modify the experimental set-up as follows.  If we delay photon $E$ with respect to photon $D$, which can be accomplished using a translation stage, then the two-photon interference at the beam splitter no longer occurs. The two pathways that lead to a coincidence measurement at $B$ and $F$ remain the same, but they no longer act coherently. The gate can instead be understood to project into the symmetric and anti-symmetric subspaces: in terms of the operators $\mathbb{S}$ and $\mathbb{A}$ defined in supplementary equation~\ref{eq:projectors},
  \begin{align}
    \mathcal{E}_{BF|DE}(\cdot)=\mathbb{S}(\cdot)\mathbb{S}+\mathbb{A}(\cdot)\mathbb{A}.
  \end{align}
Substituting the expressions from supplementary equation~\ref{eq:projectors}, we find that  this expression is equivalent to the probabilistic mixture of identity and swap of supplementary equation~\ref{eq:ProQ gate}:
  \begin{align}
    \begin{split}
      \mathcal{E}_{BF|DE}(\cdot)&=\frac{1}{2}(\mathds{1}_{B|D} \otimes
      \mathds{1}_{F|E})~(\cdot)~\left(\mathds{1}_{B|D}
     \otimes \mathds{1}_{F|E}\right)\\
     &\hspace{1cm}+ \frac{1}{2}(\mathds{1}_{B|E} \otimes \mathds{1}_{F|D})~(\cdot)~\left(\mathds{1}_{B|E} \otimes
     \mathds{1}_{F|D}\right).
  \end{split}
  \end{align}
  We note that, unlike the previous case, the overall gate does not depend on the
  relative phase $\theta$ of the clockwise and anti-clockwise paths through the
  Sagnac interferometer. Here, the beam splitter reflectivity adjusts the relative
  weights of identity and swap. For a 50-50 beam splitter, as used in the experiment,
  both the identity and swap operations will have equal weights and we obtain the
  circuit of Fig.~\ref{fig:ExampleCircuits}(c), which implements an example of the class  \textsc{ProbQ}.  

   \subsubsection*{Example of \textsc{ProbC}}
   In order to experimentally realize our example of the class \textsc{ProbC}, we
   modify the gate $\mathcal{E}_{BF|DE}$ from supplementary equation~\ref{eq:cohgate} by applying complete dephasing along the $\hat z$ axis to $E$, $D$, and $B$, so that
 \begin{align}\label{eq:channel}
   \begin{split}
      &\mathcal{E}_{BF|DE}(\cdot)=\\ 
      &(\Delta^B_{\hat z}\otimes\mathcal{I}_F)\left(U_{BF|DE}\left( (
      \Delta^D_{\hat z} \otimes\Delta^E_{\hat{z}})(\cdot)\right)
      U_{BF|DE}^{\dag}\right).
    \end{split}
 \end{align}
 Using supplementary equation~\ref{eq:general causal map}, the causal map is found to be  
 \begin{align} 
   \mathcal{E}_{CB|D}(\cdot)=&\frac{1}{2}\tfrac{1}{2}\id_C\otimes\Delta^B_{\hat
     z}\circ \mathcal{I}_{B|D}(\cdot)  \\ \nonumber 
     & +\frac{1}{2}\left(\Delta^C_{\hat z}\otimes\Delta^B_{\hat
     z}\right)\left(\ket{\Phi^+}\bra{\Phi^+}_{CB}\right){\rm Tr}_D\left(\cdot
     \right).
 \end{align}
One can see that this causal map takes the effectively classical form of supplementary equation~\ref{eq:classical map}, with $\ket{c}$, $\ket{b}$ and $\ket{d}$ all denoting eigenstates of $\sigma_z$, and
 \begin{align}
   P(cb|d)=\frac{1}{2}u(c)\delta_{b,d}+\frac{1}{2}\delta_{c,b}u(c),
 \end{align}
which is exactly the same as in our example of Fig.~\ref{fig:ExampleCircuits}(a). As we pointed out in the previous discussion of this example (around supplementary equation~\ref{eq:ProbC}), the causal map is classical in both pathways (due to the dephasing) and manifestly takes the form of a probabilistic mixture, hence it belongs to  \textsc{ProbC}.

  \subsubsection*{Example of \textsc{PhysC}}

 The class \textsc{PhysC} is experimentally realized by applying complete dephasing
 in the eigenbases of $\sigma_x$ on $E$, $\sigma_y$ on $D$ and
 $\sigma_z$ on $B$. This choice of bases ensures that the witness of physical
 mixture, which is evaluated using only measurements of these particular observables,
 remains unchanged by the dephasing. It therefore ensures that we continue to realize
 a physical mixture while eliminating the coherence in the cause-effect and
 common-cause paths.  Combining the dephasing channels with the partial swap unitary,
 we can write the overall two-qubit gate in this scenario as 
    \begin{align}
      \begin{split}
      &\mathcal{E}_{BF|DE}(\cdot)=\\ 
      &(\Delta^B_{\hat z}\otimes\mathcal{I}_F)\left(U_{BF|DE}\left( (
      \Delta^D_{\hat y }\otimes\Delta^E_{\hat x})(\cdot)\right)
      U_{BF|DE}^{\dag}\right).
    \end{split}
    \label{eq:gatephysC}
    \end{align}
 Since the dephasing introduces a different preferred basis for each of the qubits,
 we will use the notation $\ket{c_x}$, $\ket{d_y}$ and $\ket{b_z}$,
 with $\{c,d,b\}\in \pm1$, for the eigenstates of the Pauli operators $\sigma_x$,
 $\sigma_y$ and $\sigma_z$, respectively.
 Inserting $\mathcal{E}_{BF|DE}$  into supplementary equation~\ref{eq:general causal map}, one can
 obtain the causal map $\mathcal{E}_{CB|D}$, which takes the special form shown in
 supplementary equation~\ref{eq:classical map} with $P(cb|d)$ given by supplementary equation~\ref{eq:PPhysC}. Similarly, the corresponding Choi state is diagonal in
 the bases $\ket{\pm_x}$ on $C$, $\ket{\pm_z}$ on $B$  and $\ket{\pm_y}$ on $D$:
  \begin{align}\label{eq:tauPhysC}
      \tau_{CBD}= &\tfrac{1}{16}\id_{CBD}\\ \nonumber
      &+\tfrac{1}{8}\ket{+_x}\bra{+_x}_C\otimes{\ket{-_z}\bra{-_z}_B}\otimes
     \ket{-_y}\bra{-_y}_D\\ \nonumber
      &+\tfrac{1}{8}\ket{-_x}\bra{-_x}_C\otimes{\ket{-_z}\bra{-_z}_B}\otimes
      \ket{+_y}\bra{+_y}_D \\ \nonumber
      &+\tfrac{1}{8}\ket{+_x}\bra{+_x}_C\otimes{\ket{+_z}\bra{+_z}_B}\otimes
      \ket{+_y}\bra{+_y}_D\\ \nonumber
      &+\tfrac{1}{8}\ket{-_x}\bra{-_x}_C\otimes{\ket{+_z}\bra{+_z}_B}\otimes
      \ket{-_y}\bra{-_y}_D  
    \end{align}
(As pointed out after supplementary equation~\ref{eq:classical choi}, the basis $\tilde{\ket{d}}$ that diagonalizes the Choi state is related to the basis $\ket{d}$ that diagonalizes the causal map by complex conjugation in the basis that defines the Choi isomorphism. In our case, $\ket{d}$ are eigenstates of $\sigma_y$, and the Choi isomorphism is defined by the eigenbasis of $\sigma_z$, hence $\tilde{\ket{d}}$ are also eigenstates of $\sigma_y$, albeit with the opposite eigenvalues.)
 
Since the output $B$ depends on whether $C$ and $D$ are correlated or anti-correlated, the causal structure cannot be described by a probabilistic mixture of purely cause-effect and purely common-cause mechanisms. Indeed, one can see that the causal map is effectively classical, since it takes the form of supplementary equation~\ref{eq:classical map}, and the classical conditional distribution $P(cb|d)$ has the same form as supplementary equation~\ref{eq:PPhysC} (up to an exchange of positive and negative correlations). 
Since both pathways of the causal map are classical, we conclude that it belongs to \textsc{PhysC}.

  \subsection*{Supplementary Note 4: Reconstructing the causal map and obtaining the
  negativity from experimental data}
 This section details how we reconstruct causal maps from experimental data using
 a maximum likelihood estimation.  The analysis is based on \cite{RiedEtAl_2015}.   

 The measurement statistics obtained in the experiment take the form of count numbers for different combinations of wave plate orientations. We measure the Pauli observable $\sigma_s$ on $C$ and $\sigma_u$ on $B$, denoting the resulting eigenvalues by $c$ and $b$, respectively, and prepare the $d$ eigenstate of $\sigma_t$ on $D$, where $s,t,u\in\{1,2,3\}$ range over $\sigma_1\equiv \sigma_x$, $\sigma_2 \equiv \sigma_y$ and $\sigma_3 \equiv \sigma_z$. Since the orientation of the wave plates encodes both the choice of observable and the selected eigenstate, the outcome in this case is not one of two possible eigenvalues, but rather whether the photon reaches the detector in the end, indicating that it was in the desired eigenstate. 
 The observed count numbers for the wave plate orientations specified by $cbdstu$ are denoted $ \tilde P^{\rm obs}(cbdstu)$.  The expected count numbers for wave plate orientations encoding $s,c$, $t,d$ and $u,b$ are therefore proportional to the joint probabilities of realizing the eigenvalues $c,b,d$ and the choices of Pauli operators $s,t,u$. We denote the expected count numbers predicted by the fitting model by
  \begin{align}
    \tilde P^{\rm fit}(cbdstu)=N P^{\rm fit}(cbdstu),
  \end{align} 
 where $P^{\rm fit}(cbdstu)$ is the joint probability distribution predicted by the
 fitting model and N is the number of runs of the experiment.

 Let us now relate the joint probability distribution $P^{\rm fit}(cbdstu)$ to the model parameters, in particular the Choi state $\tau_{CBD}$ which represents the causal map. To this end, we introduce the notation $\Pi^{s,c}$ for the projector onto the $c\in\{\pm1\}$ eigenstate of the Pauli operator $\sigma_s$, with $s\in\{1,2,3\}$.  The conditional probability of finding eigenvalues $c,b$, given that one chose Pauli observables $s,t,u$ and prepared the $d$ eigenstate on $D$, can then be written in terms of the causal map and its Choi state as
 \begin{align} 
\label{CondProb}
P^{\rm fit}(cb|dstu) &\equiv {\rm Tr} \left[  \Pi^{s,c}_{C}\otimes\Pi^{u,b}_{B}
    \mathcal{E}_{CB|D} \left(\Pi^{t,d}_{D}\right) \right]\\ \nonumber &\equiv 2 {\rm
    Tr} \left[ T_D\left( \tau_{CBD}\right) \Pi^{s,c}_{C}\otimes\Pi^{u,b}_{B} \otimes
    \Pi^{t,d}_{D}\right],
  \end{align}
where $T_D$ denotes the  transpose with respect to the input system $D$.
In our experiment, we choose which eigenstate $d$ to prepare by rotating the wave plates after the polarizing beam-splitter, with each setting being implemented for an equal period of time. Under the assumptions of a constant rate of photon production (on average) and equal transmission efficiency of the wave plates with different settings, this can be modelled by simply taking the probability of each eigenvalue to be $P(d|t)=\frac{1}{2}$ for $d=\pm1$, $t=1,2,3$.
In this case, the probabilities of eigenvalues $c,b,d$ given the settings (choices of eigenbasis) $s,t,u$ become
\begin{align}
 \label{eq:Pklm}
    P^{\rm fit}(cbd|stu) &=P^{\rm fit}(cb|dstu)P^{\rm fit}(d|t)\\
  \nonumber&={\rm Tr} \left[ \tau_{CBD}\cdot
  \Pi^{s,c}_{C}\otimes\Pi^{u,b}_{B} \otimes T_D\left( \Pi^{t,d}_{D} \right)\right],
  \end{align}
where we note that the probability of outcome $d$ in the measurement on $D$ is independent of the settings of the other two measurements, that is, $P(d|stu)=P(d|t)$. 
Finally, we note that the choice of observables $s,t,u$ in our experiment is made at random, so that $P(stu)=\frac{1}{27}$ for all values of $s,t,u$. Using the chain rule $P(cbdstu)=P(cbd|stu)P(stu)$, we can finally write the expected count numbers in terms of the model parameters: 
 \begin{align}
\label{JointCounts}
    \tilde P^{\rm fit}(cbdstu)= {\rm Tr} \left[ \frac{N}{27}\tau_{CBD}\cdot
  \Pi^{s,c}_{C}\otimes\Pi^{u,b}_{B} \otimes T_D\left( \Pi^{t,d}_{D} \right)\right].
  \end{align} 

The operator $\tau_{CBD}$ that parametrizes the model is subject to certain
consistency constraints: as a Choi state, it must be positive semi-definite and have
trace one, while the combination $\frac{N}{27}\tau_{CBD}$ need only be positive, but
not normalized. Following Ref.~[\onlinecite{james_measurement_2001}], this is achieved with the following parameterization:
\begin{align} 
  \tau_{CBD}=\frac{N}{27}J^{\dag}_{CBD}J_{CBD},
  \end{align}
where $J_{CBD}$ is an 8x8 lower triangular matrix with real diagonal elements, specified by 64 real parameters. This form, known as the Cholesky decomposition, is positive-semidefinite by design, and, by varying over $J_{CBD}$, ranges over all positive operators. We normalize to trace one after the optimization by dividing by ${\rm Tr}(J^{\dag}_{CBD}J_{CBD})$. 

A second constraint arises due to the particular configuration of our experimental setup: since the preparation D occurs after the measurement of C, the input at D cannot have any causal influence on the measurement outcome at C. Therefore, the marginal $\tau_{CD}\equiv \mathrm{Tr_B}(\tau_{CBD})$ must be independent of $D$,
  \begin{align}
  \label{eq:uniform marginal}
   \tau_{CD}=\rho_C\otimes\frac{\id}{2},
  \end{align}
  where $\rho_C={\rm Tr_{BD}}(\tau_{CBD})$. 

We include this additional constraint in the least-squares fit by adding penalty functions to the residue, so that the overall argument becomes
  \begin{align}\label{eq:general reconstruction} 
   \begin{split}
   \chi^2=&\sum_{cbdstu} \frac{\left[ \tilde{P}^{\rm obs}(cbdstu) - \tilde{P}^{\rm
   fit}(cbdstu) \right] ^2}{\tilde{P}^{\rm fit}(cbdstu)}\\
   &+\lambda\sum_{ij}\left|(\tau_{CD}-\rho_C\otimes\frac{\id}{2})_{ij}\right|^2.  
 \end{split}
  \end{align}
  To enforce these constraints but not overshadow the principal function, the value of the Lagrange multiplier, $\lambda$ was selected heuristically to be $10^7$.

\subsubsection*{Obtaining the negativity from experimental data}

 We now describe how to obtain the negativity of the induced states $\tau^d_{CB}$, $\tau^c_{BD}$, and $\tau^b_{CD}$ from experimental data. Since the method is similar for all three, we will only illustrate it for the last case. In order to measure entanglement in $\tau^b_{CD}$, we first reconstruct the state as a 2-qubit operator on $C$ and $D$, using only count numbers from runs in which we found a particular eigenstate $\Pi^{u,b}$ on $B$, denoted $\tilde P^{\rm obs}(cdst|bu)$.  Following the model from the least-squares reconstruction of the full causal map, and still assuming that the preparations on $D$ can be modelled with $P(d)=\frac{1}{2}$, the joint count numbers take the form of supplementary equation~\ref{JointCounts}. 
 We post-select on an outcome $b$, assuming the measurement basis $u$ to be fixed.
 This gives rise to the conditional distribution 
\begin{align}
 \label{eq:pfit induced choi}
 \tilde P^{\rm fit}&(cdst|bu)=\\\nonumber
 &=\frac{N}{9} {\rm Tr}_{CD} \left[ \frac{1}{P(b|u)} {\rm Tr}_B \left( \Pi^{u,b}_{B}
 \tau_{CBD} \right)   \Pi^{s,c}_{C} \otimes T_D\left( \Pi^{t,d}_{D}
 \right)\right]\\\nonumber
 &=\frac{N}{9} {\rm Tr}_{CD} \left[\tau_{CD}^b ~ ~ \Pi^{s,c}_{C} \otimes T_D\left( \Pi^{t,d}_{D} \right)\right].
\end{align} 
 Although a full specification of the state on $B$ on which we post-select specifies both the eigenvalue $b$ and the choice of observable $u$, since the latter is assumed fixed, we suppress it and write simply  $\tau_{CD}^{b}$. The model against which we compare the observed count numbers can therefore be written as
 \begin{eqnarray}
 \tilde P^{\rm fit}(cdst|b)
 ={\rm Tr}_{CD} \left[ \frac{N}{9} \tau_{CD}^{b}  \Pi^{s,c}_{C} \otimes T_D\left( \Pi^{t,d}_{D} \right)\right].
\end{eqnarray} 

 The reconstruction method is then essentially the same as for the full causal maps: we parametrize $\tau^{b}_{CD}$ as a 4x4 lower triangular matrix with 16 real parameters and minimize the residue 
 \begin{align} 
   \begin{split}
   \chi^2=&\sum_{cdst} \frac{\left[ \tilde{P}^{\rm obs}(cdst) - \tilde{P}^{\rm
   fit}(cdst) \right] ^2}{\tilde{P}^{\rm fit}(cdst)}.
 \end{split}
  \end{align} 
 Once the optimal parameters have been found, the negativity, $\mathcal{N}$, of the
 reconstructed state $\tau^{b}_{CD}$ is calculated using equation~\ref{eq:negativity}. Negativities for the induced states $\tau^d_{CB}$ and $\tau^c_{BD}$ are calculated in a similar way.

\subsection*{Supplementary Note 5: Witness of physical mixture}
  In this section, we define a family of functions of the experimental statistics that witness physical mixtures of common-cause and cause-effect mechanisms. In other words, we seek functions that are zero for all probabilistic mixtures and non-zero for at least some physical  mixtures. For simplicity, we restrict ourselves to the case of qubits.

  The witness is defined in terms of the statistics of measurements of a single Pauli  observable on each $C$ and $B$ and preparations of eigenstates of a third Pauli  observable on $D$. That is, the settings $s,t,u$ are fixed, with each choice giving  rise to a different witness from the same family, and we omit them in the  following for brevity. When calculating the witness, we choose the eigenvalue $d$  for the preparation of $D$ from the uniform distribution, $P(d)=\frac{1}{2}$ for  $d=\pm1$, and hence the joint probability distribution $P(cdb)$ takes the same form as in  supplementary equation~\ref{eq:Pklm}.

\subsubsection*{Properties of probabilistic mixtures}

We begin by noting several mathematical properties of probabilistic mixtures that will be useful in the subsequent derivations. 

As already shown, the Choi state representing a causal map that is a probabilistic
mixture of common-cause and cause-effect can always be expressed as a sum of only two
terms,
\begin{equation}
  \label{probmixrho} \tau^{prob}_{CBD}=p \rho_{CB} \otimes \frac{1}{2}\mathbb{1}_{D}
  +(1-p)  \rho_C \otimes  \tau_{BD}. 
\end{equation}
The first term represents the common-cause scenario, wherein we prepare a bipartite state $\rho_{CB}$ and trace out $D$; hence the marginal on $D$ of the Choi state is the completely mixed state. The state $\rho_{CB}$ is obtained from the initial state $\rho_{CE}$ by a CPTP map that takes $E$ to $B$, hence the marginal on $C$ is unchanged: ${\rm Tr}_B \rho_{CB}={\rm Tr}_E \rho_{CE}$. The second term corresponds to a cause-effect scenario, in which case the marginal state on $C$ is simply the marginal of the initial bipartite state $\rho_{CE}$, $\rho_C={\rm Tr}_E \rho_{CE}$. Meanwhile, $\tau_{BD}$ is the Choi state corresponding to a CPTP map from $D$ to $B$, hence its marginal on $D$ is again the completely mixed state. In summary, the marginals of the two terms on $C$ and $D$, respectively, are equal:
\begin{align}
&{\rm Tr}_B \rho_{CB}={\rm Tr}_E \rho_{CE}=\rho_C,\\
&{\rm Tr}_B  \tau_{BD}=\frac{1}{2}\mathbb{1}_{D}.
\end{align}
It furthermore holds for all causal maps that $C$ and $D$ become independent if we ignore $B$:
\begin{equation}
{\rm Tr}_B \left[\tau_{CBD}\right]=\rho_{C} \otimes \frac{1}{2}\mathbb{1}_{D}.
\end{equation}

The experimental statistics inherit these properties: letting $u(d)\equiv \frac{1}{2}$ $\forall d=\pm1$ denote the uniform probability distribution, we have
\begin{align}
\label{probmixstatistics}
P(cdb)&=p P_{CB}(cb)u(d)+(1-p)P_C(c) P_{BD}(bd).
\end{align}
The marginal distributions over $c$ and $d$ in both terms are identical,
\begin{align}
\sum_b P_{CB}(cb)=P_C(c),\\
\sum_b P_{BD}(bd)=u(d),
\end{align}
and, if we ignore $b$, then $c$ and $d$ become independent,
\begin{equation}
\label{margindep}
\sum_b P(cdb)=P_C(c)u(d).
\end{equation}

\subsubsection*{Intuitive simple version of the witness}

Suppose that the marginal on $C$ is completely mixed, so that a probabilistic mixture of common-cause and cause-effect takes the form
\begin{align}
\tau'_{CBD}=p \rho_{CB} \otimes \frac{1}{2}\mathbb{1}_{D} +(1-p)  \frac{1}{2}\mathbb{1}_{C} \otimes  \tau_{BD},
\label{eq:probmixt}
\end{align}
with
\begin{equation}
{\rm Tr}_B \rho_{CB}=\frac{1}{2}\mathbb{1}_{C},
~~
{\rm Tr}_B  \tau_{BD}=\frac{1}{2}\mathbb{1}_{D}.
\end{equation}
Under this assumption, we can construct a witness in terms of the joint probabilities of supplementary equation~\ref{eq:Pklm} which is simply the expectation value of $\tau_{CBD}$ for a product of Pauli observables, 
\begin{align}
  \begin{split}
  \mathcal{C}_{CD}^0& \equiv \sum_{cdb}cdb P(cdb)\\
&={\rm Tr}\left[\tau_{CBD}\sigma^s_C\otimes\sigma^u_B\otimes T_D\left( \sigma^t_D \right)\right],
\end{split}
\label{eq:W0}
\end{align}
For any ${s,t,u}\in{1,2,3}$, this is zero for any probabilistic mixture, as can be seen by
inserting supplementary equation~\ref{eq:probmixt} into supplementary equation~\ref{eq:W0}. Therefore, if one can assume that
$\rho_C=\frac{1}{2}\mathbb{1}$, then non-zero value
of $\mathcal{C}_{CD}^0$ heralds a physical mixture.

\subsubsection*{General form of the witness}

If we cannot justify the assumption that $\rho_C=\frac{1}{2}\mathbb{1}$, then we must use a more general version of the witness. We will now propose such a witness: a measure of induced correlations that is designed to be zero for probabilistic mixtures even if $\rho_C\neq\frac{1}{2}\mathbb{1}$.

Given the joint distribution $P(cbd)$, one can calculate the marginal $P(b)=\sum_{cd} P(cbd)$ and the conditional distribution $P(cd|b)=P(cbd)/P(b)$. For each value of $b$, the latter is a distribution over $c$ and $d$, and therefore the correlations between the two can be quantified by their {covariance},
\begin{align}
{\rm cov}(c,d|b)& =\sum_{cd}cd P(cd|b)  \\\nonumber 
& -\left[\sum_{cd} c P(cd|b)\right] \left[\sum_{cd} d P(cd|b)\right].
\end{align}
We now define our witness to be  the weighted difference of the covariances in the conditional distributions,
\begin{equation}
  \mathcal{C}_{CD}=2\sum_{b=\pm1} b P(b)^2 {\rm cov}(cd|b).
\end{equation}
We will prove that this choice has the desired properties in the following.

\subsubsection*{Simplification in limiting case}
The witness $\mathcal{C}_{CD}$ reduces to $\mathcal{C}_{CD}^0$ if certain marginals of $P(cdb)$ are uniform, specifically, if
\begin{align}
P(cb)\equiv \sum_d P(cdb)=\frac{1}{4} ~\forall c,b,\\
P(db)\equiv \sum_c P(cdb)=\frac{1}{4} ~\forall d,b.
\end{align}
This ensures that each $b$ occurs with equal probability, $P(b)=\frac{1}{2}$, and consequently the conditional distributions also satisfy
\begin{align}
P(c|b)\equiv \sum_d P(cd|b)=\frac{1}{2} ~\forall c,b,\\
P(d|b)\equiv \sum_c P(cd|b)=\frac{1}{2} ~\forall d,b,
\end{align}
that is, the conditional distribution $P(cd|b)$ has uniform marginals on $c$ and $d$. The expectation values $\left \langle c \right \rangle$ and $\left \langle d \right \rangle$ under this distribution are zero, so that the covariance simplifies to 
\begin{equation}
{\rm cov}(c,d|b) =\left \langle cd \right \rangle= \sum_{cd} cd P(cd|b),
\end{equation}
and therefore
\begin{align}\nonumber
  \mathcal{C}_{CD} &= 2\sum_b b P(b)^2 \sum_{cd} cd P(cd|b) \\
  &=\sum_{cdb} cdb P(cdb)\equiv \mathcal{C}_{CD}^0.
\end{align}
In this sense, the witness $\mathcal{C}_{CD}$ is a generalization of the expectation value of the simple product of Paulis that defines $\mathcal{C}_{CD}^0$

\subsubsection*{Casting $\mathcal{C}_{CD}=0$ directly in terms of count numbers}

In order to facilitate the proof below as well as the assessment of whether or not $\mathcal{C}_{CD}=0$ based on experimental data, we cast the witness in a different form. To this end, we note that, if $c$ and $d$ are binary variables whose values are labelled $\pm1$, then their covariance under a conditional distribution $P(cd|b)$ takes the form
\begin{equation}
{\rm cov}(c,d|b)=4 \left[P(++|b) P(--|b) - P(+-|b) P(-+|b)\right].
\end{equation}
This allows us to rewrite the witness in terms of the joint probabilities $P(cbd)$ as
\begin{equation}
\label{witness}
\mathcal{C}_{CD} \equiv 8 \sum_{b=\pm1} b \left[  P(++b) P(--b) - P(+-b) P(-+b)  \right].
\end{equation}

\subsubsection*{$\mathcal{C}_{CD}=0$ for probabilistic mixtures}

Now we can show that $\mathcal{C}_{CD}$ is zero for any probabilistic mixture of common-cause and cause-effect relations. Recall that, since $b$ only takes two values, the marginal independence (supplementary equation~\ref{margindep}),
\begin{equation}
\sum_b P(cdb)=P_C(c)u(d)=\frac{1}{2}P_C(c)
\end{equation}
 implies that
 \begin{equation}
 P(cd,-)=\frac{P_C(c)}{2}-P(cd,+).
 \end{equation}
This allows us to rewrite the $b=-1$ term in supplementary equation~\ref{witness} as
\begin{align}
  \begin{split}
&P(++,-) P(--,-) - P(+-,-) P(-+,-)\\
&=-\frac{P_C(+)}{2}P(--,+)-\frac{P_C(-)}{2}P(++,+)\\
&+\frac{P_C(+)}{2}P(-+,+)+\frac{P_C(-)}{2}P(+-,+)\\
&+[P(++,+) P(--,+) - P(+-,+) P(-+,+)],
\end{split}
\end{align}
hence the witness reduces to 
\begin{align}
  \begin{split}
    \mathcal{C}_{CD} &= 4 [P_C(-)P(++,+) -P_C(-)P(+-,+)\\
&-P_C(+)P(-+,+)+P_C(+)P(--,+)]
\end{split}\\
&=4\sum_{cd} cd \left[ 1-P_C(c) \right] P(cd,+)
\end{align}
Our core hypothesis, of a probabilistic mixture, implies that $P(cd,+)$ is a convex combination of two terms, each one a product distribution over $cd$. Substituting supplementary equation~\ref{probmixstatistics} and distributing the sums,
\begin{align}
  \begin{split}
    \mathcal{C}_{CD}&= 4p \left[\sum_{c} c \left[ 1-P_C(c) \right] P_{CB}(c,+)\right]\left[\sum_d d\; u(d)\right] \\
&+4(1-p)\left[ \sum_c c \left[ 1-P_C(c) \right] P_C(c)\right] \left[\sum_d  P_{BD}(d,+)\right].
\end{split}
\end{align}
In the first term, we have the average over $d=\pm1$ under the uniform distribution, which is zero. In the second term, the sum over $c$ gives $P_C(+)P_C(-)-P_C(-)P_C(+)=0$. Thus, for any causal map that is a probabilistic mixture of cause-effect and common-cause mechanisms of the form of supplementary equation~\ref{probmixrho}, we have
\begin{equation}
  \mathcal{C}_{CD}=0.
\end{equation}

\subsubsection*{Measuring the witness from experimental data}
  We calculate the witness $\mathcal{C}_{CD}$ explicitly from experimental count numbers $\tilde P(c,d,b)$
  using supplementary equation~\ref{witness},
  \begin{align}
    \mathcal{C}_{CD}=\frac{\displaystyle\sum_{b=\pm1}b\left( \tilde P(++b)\tilde
    P(--b)-\tilde P(+-b)\tilde P(-+b)\right)}{\left(\displaystyle\sum_{c,d,b=\pm1}\tilde P(c,d,b)\right)^2}.
  \label{eq:witness counts} \end{align}
 The uncertainty on the witness is calculated by assuming Poissonian noise on the count numbers and propagating the errors through supplementary equation~\ref{eq:witness counts}. 

\subsection*{Supplementary Note 6: Bounds on induced mutual information in Berkson's paradox}

In the following, we derive an upper bound on the mutual information between two causes, $D$ and $E$, conditioned on their common effect, $B$, under the assumption that the two influences are combined probabilistically, that is,
\begin{equation}
P(B|DE)=(1-p)P_{\mathcal{D}}(B|D)+pP_{\mathcal{E}}(B|E).
\end{equation}
The derivation is cast in terms of classical variables, but an extension to the quantum case is given at the end. 

The distribution over $DE$ conditional on some value of $B$ can be obtained by Bayesian inversion. 
Note that, since $D$ and $E$ do not share a common cause, our prior probability distribution over them takes the form of a product of two generic probability distributions, which we denote by $Q(D)$ and $Q(E)$. 
It follows that
\begin{align}
  \begin{split}
P(DE|B)&\equiv P(B|DE) P(DE) /P(B)\\
&=(1-p) Q(E) \frac{P_{\mathcal{D}}(B|D)Q(D)}{P(B)}\\
& + p~ Q(D) \frac{P_{\mathcal{E}}(B|E) Q(E)}{P(B)},
\end{split}
\end{align}
where $P(B)\equiv\sum_{DE} P(B|DE)Q(D)Q(E)$.
For each value $b$ of $B$, the fractions are distributions over $D$ and $E$, respectively, but not necessarily normalized. Let $P^b_{\mathcal{D}}(D)$ and $P^b_{\mathcal{E}}(E)$ denote the corresponding normalized distributions, introducing the $b$-dependent modified weight $q^b$ to absorb the difference in normalization:
\begin{equation}
q^b=p\frac{1}{P(B=b)}\sum_E P_{\mathcal{E}}(B=b|E)Q(E),
\end{equation}
or, equivalently,
\begin{equation}
(1-q^b)=(1-p)\frac{1}{P(B=b)}\sum_D P_{\mathcal{D}}(B=b|D)Q(D),
\end{equation}
and
\begin{align}
&P^b_{\mathcal{E}}(E)=\frac{p}{q^b} \frac{1}{P(B=b)}P_{\mathcal{E}}(B=b|E)Q(E)\\
&P^b_{\mathcal{D}}(D)=\frac{1-p}{1-q^b}\frac{1}{P(B=b)} P_{\mathcal{D}}(B=b|D)Q(D).
\end{align}
For the purpose of this derivation, we will focus on a single value $b$ and, for brevity, suppress the explicit $b$-dependence in the following.
In this new notation,
\begin{align}
\label{DEprobmix}
P(DE)=(1-q) Q(E) P_{\mathcal{D}}(D)+ q Q(D) P_{\mathcal{E}}(E).
\end{align}

We will show that the mutual information $I(D:E)$ in this distribution is maximal if $P_{\mathcal{D}}(D)$ and $P_{\mathcal{E}}(E)$ each produce a single value with certainty. 
To see this, consider the mutual information as a functional of two arguments, the marginal distribution over $E$,
\begin{equation}
P(E)=(1-q)Q(E)+qP_{\mathcal{E}}(E),
\end{equation}
and the conditional 
\begin{align}
  \begin{split}
P(D|E)&=\frac{(1-q)Q(E)}{(1-q)Q(E)+qP_{\mathcal{E}}(E)} P_{\mathcal{D}}(D) \\
&+\frac{qP_{\mathcal{E}}(E)}{(1-q)Q(E)+qP_E(E)} Q(D).
\end{split}
\end{align}
One can show (Ref.~\cite{CoverThomas_book}, theorem 2.7.4) that the mutual information is convex in the second argument, that is, for a fixed marginal $P(E)$, 
 \begin{align}
   \begin{split}
 I(D:E)&\left[P(E),\lambda P^0(D|E)+(1-\lambda) P^1(D|E)\right] \\
&\le\lambda I(D:E)\left[P(E),P^0(D|E)\right] \\
&+(1-\lambda)I(D:E)\left[P(E),P^1(D|E)\right].
\end{split}
 \end{align}
In order to apply this fact to our problem, suppose that we fix the marginal $P(E)$ -- and consequently the fractions in the expression for $P(D|E)$ above -- but take a convex combination 
\begin{equation}
P_{\mathcal{D}}(D)=\lambda P_{\mathcal{D}}^0(D)+(1-\lambda)P_{\mathcal{D}}^1(D),
\end{equation}
so that the resulting $P(D|E)$ is a convex combination with weight $\lambda$ as well. In this case, an upper bound on the mutual information follows. It follows that, for fixed $P_{\mathcal{E}}(E)$ and $q$, the largest mutual information is achieved when the distribution $P_{\mathcal{D}}(D)$ is extremal, meaning that it produces one value with certainty. We express this as $P_{\mathcal{D}}(D)=\delta(D)$ for short. We do not specify which value of $D$ is found with certainty, since the mutual information depends only on the probabilities of different values, but not on their labels. By symmetry, in order to maximize the mutual information we must also have $P_{\mathcal{E}}(E)=\delta(E)$.

The maximal mutual information between $D$ and $E$ for a distribution constrained to the form (\ref{DEprobmix}) is therefore achieved by a distribution of the form
\begin{align}
P(DE)=(1-q) Q(E) \delta(D)+ q Q(D) \delta(E).
\end{align}
In order to evaluate the maximal mutual information explicitly, we make two simplifying assumptions: first, let us assume that the prior distributions $Q(D)$ and $Q(E)$ are both uniform, that is, that we have no additional information about them beyond what we can retrodict from $B$. Let us furthermore assume
that $D$ and $E$ range over an equal number of values, $N$. Symmetry then suggests that the mutual information is maximal when $q=\frac{1}{2}$, which can be verified analytically. 
In this case, we obtain
\begin{equation}
I(D:E)=\log N - \frac{N+1}{N} \left[ \log (N+1)-1\right]
\end{equation}
with $\log$ denoting the logarithm used to calculate the entropy. By contrast, the maximal mutual information between $D$ and $E$ without any constraints is $\log N$. If $D$ and $E$ are bits ($N=2$) and we calculate the logarithms in base 2, the upper bound on the mutual information becomes
\begin{equation}
I(D:E)\le \frac{5}{2}-\frac{3}{2}\log_2(3) \approx .12.
\end{equation}

Now consider the case where $D$ and $E$ are quantum systems. Their state under post-selection on a measurement outcome $b$ on $B$ can be written
\begin{equation}
\rho_{DE}^{b}=(1-q)\rho_D\otimes\frac{\mathbb{1}_{E}}{2}+q\frac{\mathbb{1}_{D}}{2}\otimes\rho_{E}.
\end{equation}
As in the classical case, we consider the prior over $D$ and $E$ to be uniform, that is, the maximally mixed quantum state. This implies that there is in fact only one non-trivial density operator on $D$ (in the first term) and $E$ (in the second) in the entire problem. Consequently there exist preferred bases of $\mathcal{H}_D$ and $\mathcal{H}_E$, namely the eigenbases of $\rho_D$ and $\rho_E$, in which all density operators of interest are diagonal and thus effectively reduced to classical probability distributions. 
Therefore the results from the classical case carry over, and we recover the upper bound above as a function of the dimension of the Hilbert spaces $N=\dim\mathcal{H}_{D}=\dim\mathcal{H}_{E}$.

\section*{Supplementary Discussion}
\subsection*{Supplementary Discussion 1: Related work on superpositions of causal orders}

We here discuss related work that considers the question of whether one can prepare a quantum-coherent mixture of different causal {orders}~\cite{Chiribella_2012,Oreshkov2012}.  For a pair of quantum systems, $A$ and $B$, the idea is to prepare a quantum-coherent mixture of $A$ being the cause of $B$ and of $B$ being the cause of $A$. 
By contrast, in this article we seek only to prepare a quantum-coherent mixture of $A$ being a cause of $B$ and of $A$ and $B$ having a common cause.  There is an important difference between the two objectives.  In our case, $A$ and $B$ can be time-like separated, 
with $A$ to the past of $B$.  In the case of a quantum-coherent mixture of causal orders, on the other hand, the temporal order is different in the two terms of the mixture and consequently these must be embedded differently in space-time.   This is the sense in which achieving a quantum-coherent mixture of causal {orders} requires one to 
abandon the assumption of a pre-defined global causal structure.

Nonetheless, the approach of defining probabilistic, physical, and quantum-coherent mixtures of causal relations that is espoused in the present article can be applied to the case of two cause-effect relations, in particular, $A$ causing $B$ and $B$ causing $A$, and it is interesting to see what lessons are learned from doing so.  We begin with the classical case.

In our approach, the overall causal structure of a given scenario is depicted by a directed acyclic graph (DAG).  
If one considers a probabilistic mixture of causal relations, then one must include enough causal influences in the graph to accommodate the causal relations that hold in any given element of the mixture.  The DAG associated to a probabilistic mixture of cause-effect and common-cause, depicted in \ref{fig:controlvariableJ}(b), therefore includes {both} a cause-effect pathway and a common-cause pathway between $A$ and $B$.  Similarly, it follows that the graph associated to a probabilistic mixture of $A$ causing $B$ and $B$ causing $A$ must have {both} a pathway wherein $A$ causes $B$ and another wherein $B$ causes $A$.  

\begin{figure}[h]
  \centering
  \includegraphics[scale=0.7]{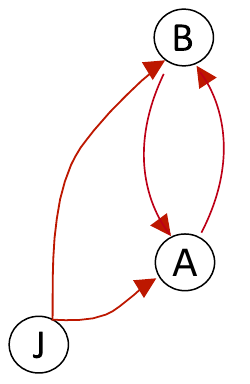}
  \caption{\textbf{Mixture of causal orders}
  In order to physically realize a probabilistic mixture of different causal orders,
  one requires a switch variable $J$ that can influence both variables $A$ and $B$.
  One also requires a cycle in the causal structure. 
  }
  \label{fig:DAGforMixOfCausalOrders}
\end{figure}

Furthermore, as noted previously, in order to physically realize a probabilistic
mixture of different causal relations, one requires a switch variable $J$ that can
influence one or more variables in the system and modify how they causally depend on
other variables. For a probabilistic mixture of cause-effect and common-cause
relations, it was shown  in Supplementary Note 1 
that this switch variable must influence $B$ alone.  But what does it imply for a probabilistic mixture of $A$ causing $B$ and $B$ causing $A$?  
In this case, the switch variable (call it $J$) cannot influence $B$ alone, but must instead influence {both} $A$ and $B$.  This is because as one varies between $J=0$ and $J=1$, 
 $A$ must toggle between having a causal dependence on $B$ and not having such a dependence while $B$ must simultaneously toggle between not having a causal dependence on $A$ and having such a dependence.  
As such, the switch variable defines a common-cause pathway between $A$ and $B$.  The overall causal structure is depicted in \ref{fig:DAGforMixOfCausalOrders}.

One can immediately observe two uncomfortable facts about the overall causal structure.  

First, the natural constraint on physical realizability of probabilistic mixtures
articulated  in Supplementary Note 1 
has been violated.  The objective was to have a probabilistic mixture of causal relations between $A$ and $B$ every element of which was purely cause-effect (either $A$ causing $B$ or $B$ causing $A$).  However, any attempt to physically realize such a mixture introduces a causal relation that is not purely cause-effect, namely, the common cause $J$.  

Second, and more importantly, one notes that the overall causal structure is not a directed {acyclic} graph because it includes a cycle.  It is unclear how to make sense of such graph.  One can no longer interpret the causal relations therein using the interventionist notion of causation that is standard for directed acyclic graphs.  The reason is as follows.  The interventionist notion of causation 
presumes that causal mechanisms in the graph are autonomous: the mechanism that describes how one variable in the graph is causally influenced by its parents can be varied independently of the mechanism that describes how any other variable in the graph is causally influenced by its parents.  But this assumption of autonomy cannot be maintained in graphs with cycles.  For instance, consider a graph having a cycle between a pair of binary variables, $A$ and $B$.  If the two causal mechanisms were autonomous, then it ought to be possible to take them to be $A := B$ and $B := A \oplus 1$ respectively.  But the latter pair of mechanisms yields a contradiction, so the mechanisms cannot be varied independently of one another.

The second of these concerns may be surmountable in the case of a {probabilistic} mixture of $A$ causing $B$ and of $B$ causing $A$, since only one of the two pathways is active for a given value of the switch variable $J$.  

If, however, one considers instead a {physical mixture} of $A$ causing $B$ and of $B$ causing $A$, then both pathways must be active simultaneously, and there is no way to deny the necessity of the cycle.

The conceptual problems introduced by the presence of cycles in the causal graph persist if one replaces classical variables with quantum systems.
In the approach we propose in this article, a quantum-coherent mixture of causal relations between quantum systems is necessarily a physical mixture of those causal relations.  Consequently, a quantum-coherent mixture of causal orders in our approach requires a graph with a cycle, with all the interpretive ambiguity that this entails.

Finally, even if one can make sense of graphs with cycles, it remains unclear how one could ever hope to realize these experimentally because in the context of relativity theory, a causal cycle is a closed time-like curve which one expects is only physically realizable in very exotic physical scenarios. 
In our approach,  therefore, realizing a quantum-coherent mixture of causal orders, if it is possible at all, is likely to only be possible in very exotic scenarios.

Some recent work by Procopio {\em et al.}~\cite{ProcopioEtAl_2015} claims to achieve an experimental realization of a superposition of causal orders in a tabletop quantum optics experiment.  
This seems to contradict our claim that one is likely to require exotic physics to achieve such a thing.  We therefore turn to the details of this experiment and 
 why we do not believe that it can be accurately described as achieving a superposition of causal orders.

The objective is to realize, in a quantum optical setting, the quantum switch proposed by Chiribella {\em et al.}~\cite{ChiribellaEtAl_2013} and explored in Ref.~\cite{Chiribella_2012}, wherein the order of two gates is controlled by an ancillary quantum system that is prepared and post-selected in a superposition of the states which prescribe a definite causal order.  This has been proposed as a means of achieving a superposition of causal orders.
The experiment is based on a folded Mach-Zehnder interferometer whereby the order of two gates, 
call them $U$ and $V$, is determined by the path taken by the photon. 
Due to the particular geometry, one requires a version of the $U$ and $V$ gate in each path of the interferometer. 

This set-up is optically equivalent to an unfolded interferometer.  In the latter case, it is clear that one requires a version of the $U$ and $V$ gate in each path of the interferometer, call them $U_1$, $V_1$ and $U_2$, $V_2$ respectively.  The different orders that one switches between are: $U_1$ is implemented and then $V_2$ is implemented, and $U_2$ is implemented and then $V_1$ is implemented.  The situation is clearly {not} one wherein one toggles between a photon passing through two fixed spatio-temporal regions in one of two different orders. 

For the case of the folded interferometer used in the experiment, it is still the case that one requires two versions of each gate; it is simply that the two versions correspond to the gate functioning {at different times}.  Call the early versions of the two gates $U_1$, $V_1$ and the late versions $U_2$, $V_2$.  Again, the different orders that one switches between are: $U_1$ is implemented and then $V_2$ is implemented, and $U_2$ is implemented and then $V_1$ is implemented. 

If one instead required that each gate act only once in a localized spatio-temporal region -- for instance, by putting shutters that let a photon through the gate only in a narrow window of time -- then the experimental set-up in question would no longer realize a quantum switch.

\end{document}